\documentclass[journal,twoside]{IEEEtran}
\usepackage{amsfonts,bm,booktabs,cite,graphicx,hyperref,multirow,textcomp,url}
\usepackage{algorithm}
\usepackage{subfigure}
\usepackage{algpseudocode}
\usepackage{amsmath}
\usepackage{mathrsfs}
\usepackage{float}
\usepackage{color}
\usepackage{booktabs}
\usepackage{bbding}
\usepackage[flushleft]{threeparttable}

\makeatletter 
\renewcommand{\@thesubfigure}{\hskip\subfiglabelskip}
 \makeatother
\begin{document}
\title{Fine-Grained Motion Compression and Selective Temporal Fusion for Neural B-Frame Video Coding}
\author{
	Xihua Sheng,  \IEEEmembership{Member, IEEE},
    Peilin Chen,
    Meng Wang, \IEEEmembership{Member, IEEE}, 
    Li Zhang, \IEEEmembership{Senior Member, IEEE}, \\
	Shiqi Wang, \IEEEmembership{Senior Member, IEEE},
	Dapeng Oliver Wu, \IEEEmembership{Fellow, IEEE}

\thanks{
This paper was received on June 7, 2025; revised on November 18, 2025; accepted on February 20, 2026; \par
This research was supported in part by the Hong Kong Innovation and Technology Commission (ITC) grant MHP/061/23, GHP/044/21SZ, and PRP/036/24FX; in part by the General Research Fund (GRF) of the RGC of Hong Kong under Grants 11205424 and 11200323, in part by the Research Grants Council (RGC) of Hong Kong under the Early Career Scheme (ECS) 23200925, and 
in part by National Natural Science Foundation of China (NSFC)/Research Grants Council (RGC) Joint Research Scheme N\_CityU198/24.

X. Sheng, P. Chen, S. Wang, and D. Wu are with the Department of Computer Science, City University of Hong Kong, Hong Kong, China (e-mail: xihsheng@cityu.edu.hk, plchen3@cityu.edu.hk, shiqwang@cityu.edu.hk, dpwu@ieee.org).\par
M. Wang is with the School of Data Science, Lingnan University, Hong Kong, China (e-mail: mengwang7@ln.edu.hk).\par
L. Zhang is with Bytedance, San Diego CA., 92122 USA, (e-mail: lizhang.idm@bytedance.com).\par
Corresponding author: Shiqi Wang.
}
}

\markboth{IEEE Transactions on Multimedia}{Fine-Grained Motion Compression and Selective Temporal Fusion for Neural B-Frame Video Coding}

\maketitle
\begin{abstract}
With the remarkable progress in neural P-frame video coding, neural B-frame coding has recently emerged as a critical research direction. However, most existing neural B-frame codecs directly adopt P-frame coding tools without adequately addressing the unique challenges of B-frame compression, leading to suboptimal performance. To bridge this gap, we propose novel enhancements for motion compression and temporal fusion for neural B-frame coding.
First, we design a fine-grained motion compression method. This method incorporates an interactive dual-branch motion auto-encoder with per-branch adaptive quantization steps, which enables fine-grained compression of bi-directional motion vectors while accommodating their asymmetric bitrate allocation and reconstruction quality requirements. Furthermore, this method involves an interactive motion entropy model that exploits correlations between bi-directional motion latent representations by interactively leveraging partitioned latent segments as directional priors. 
Second, we propose a selective temporal fusion method that predicts bi-directional fusion weights to achieve discriminative utilization of bi-directional multi-scale temporal contexts with varying qualities. Additionally, this method introduces a hyperprior-based implicit alignment mechanism for contextual entropy modeling. By treating the hyperprior as a surrogate for the contextual latent representation, this mechanism implicitly mitigates the misalignment in the fused bi-directional temporal priors.
Extensive experiments demonstrate that our proposed codec achieves an average BD-rate reduction of approximately 10\% compared to the state-of-the-art neural B-frame codec, DCVC-B, and delivers comparable or even superior compression performance to the H.266/VVC reference software under random-access configurations.

\end{abstract}
\IEEEpeerreviewmaketitle

\section{Introduction}
The dramatic growth of video traffic has created unprecedented demands for efficient video coding technologies. To improve compression efficiency, a series of video coding standards have been developed over the past few decades, including H.264/AVC~\cite{wiegand2003overview}, H.265/HEVC~\cite{sullivan2012overview}, and H.266/VVC~\cite{bross2021overview}. While these standards have achieved remarkable success, further advancements in compression performance remain challenging.\par

Recently, neural video coding has gained significant attention~\cite{lu2020end,hu2022coarse,sheng2025prediction,Rippel_2021_ICCV,hu2020improving,lu2020content,lin2020m,hu2021fvc,agustsson2020scale,cheng2019learning,rippel2019learned,liu2021deep,liu2020neural,liu2022end,yilmaz2021end,lin2022dmvc,guo2023learning,guo2023enhanced, liu2020conditional, wang2023learned,wei2025rdvc,ma2024uncertainty,DBLP:conf/nips/MentzerTMCHLA22,li2021deep, wu2025end,ho2022canf,chen2024maskcrt,sheng2022temporal,chen2023b,sheng2025bi,ye2024deep,li2022hybrid,li2023neural,sheng2024spatial,li2024neural,chen2021nerv,wang2024ssnvc,kwan2024hinerv}. Existing schemes primarily focused on two scenarios: low-delay coding and random-access coding. In the low-delay scenario, the current frame (P-frame) is constrained to reference only previously encoded frames, ensuring minimal latency. In contrast, in the random-access scenario, the current frame (B-frame) is allowed for bi-directional referencing, trading latency for higher compression efficiency. Current research efforts concentrate on two aspects: (1) how to obtain compact yet accurate motion representation and (2) how to effectively utilize predicted temporal contexts. For the first challenge, for example, the methods such as multiframe-based motion vector prediction~\cite{lin2020m}, resolution-adaptive motion vector compression~\cite{hu2020improving}, and coarse-to-fine motion vector compression~\cite{hu2022coarse} have effectively improved the accuracy of motion-compensated prediction and the efficiency of motion compression. For the second challenge, innovations such as condition coding~\cite{li2021deep} have shown advantages over traditional residual coding by better exploiting temporal correlations. However, most existing neural B-frame coding solutions~\cite{sheng2025bi,chen2023b,ye2024deep} simply adapt these P-frame coding tools without fully considering the unique characteristics of bi-directional prediction, which limits their performance potential. \par

B-frame coding exhibits two distinctive characteristics that pose unique challenges. First, it requires estimating bi-directional motion vectors from bi-directional reference frames, leading to higher motion coding costs compared to P-frame coding.  Current schemes typically process bi-directional motion vectors by either: (1) concatenating them for joint compression using a single motion auto-encoder~\cite{sheng2025bi,chen2023b}, or (2) employing a parameter-shared motion auto-encoder for separate compression~\cite{ye2024deep}. However, these coarse-grained motion vector compression methods treat bi-directional motions in an undifferentiated manner and thus overlook two fundamental aspects of bi-directional motion characteristics: (a) the temporal correlation asymmetry between forward and backward references leads to distinct bitrate allocation needs and, consequently, different required levels of precision for the reconstruction of bi-directional motion vectors, and (b) the inherent geometric consistency between forward and backward motion vectors can be exploited through joint motion modeling to improve compression efficiency. Second, the quality of predicted bi-directional temporal contexts and temporal priors varies substantially due to the differences in temporal reference correlation and motion vector precision. Existing schemes~\cite{sheng2025bi,chen2023b,ye2024deep} fused these contexts and priors uniformly through the contextual encoder-decoder and contextual entropy model without quality discrimination, which may propagate prediction errors and ultimately degrade rate-distortion performance. These characteristics suggest that dedicated algorithmic designs for B-frame coding may be necessary to achieve further compression efficiency improvement.\par

In this paper, to reduce the bi-directional motion coding costs, we propose a fine-grained motion compression method for neural B-frame video coding. Specifically, we propose an interactive dual-branch motion auto-encoder to separately compress and reconstruct forward and backward motion vectors. This motion auto-encoder uses a cross-branch motion information interaction mechanism and per-branch adaptive quantization steps to facilitate bi-directional motion information exchange and fine-grained motion rate-distortion control. Furthermore, we propose an interactive motion entropy model that separately estimates the bitrate for each directional motion latent representation. This entropy model interactively leverages partitioned segments of bi-directional motion latent representations as direction priors, thereby effectively capturing fine-grained bi-directional motion dependencies and enhancing compression efficiency.\par
To enable discriminative utilization of bi-directional temporal contexts and temporal priors, we propose a selective temporal fusion method for neural B-frame video coding. Specifically, during contextual encoding and decoding, our contextual encoder-decoder predicts bi-directional fusion weights, which serve as discriminative cues to guide the selective fusion of multi-scale bi-directional temporal contexts.
Beyond the contextual encoder-decoder, we further introduce selective temporal prior fusion in the entropy modeling stage. By treating the hyperprior as a surrogate for the contextual latent representation, we implicitly align the fused bi-directional temporal priors, thereby effectively exploiting the correlation of temporal priors.\par
Our main contributions are summarized as follows:
\begin{itemize}
    \item We propose a fine-grained motion compression method for neural B-frame video coding, which involves an interactive dual-branch motion auto-encoder,  per-branch adaptive quantization steps, and an interactive motion entropy model to reduce bi-directional motion coding costs.

    \item We propose a selective temporal fusion method for neural B-frame video coding, which uses bi-directional fusion weights and hyperprior-based implicit alignment to enable discriminative utilization of bi-directional temporal contexts and priors in the contextual encoder-decoder and contextual entropy model.

    \item Experimental results show that our codec outperforms state-of-the-art neural B-frame codecs and achieves comparable or even superior compression performance to the H.266/VVC reference software under random-access configuration.

\end{itemize}
We organize the remainder of this paper as follows.  Section~\ref{sec:related_work} reviews existing work on neural P-frame and B-frame video coding. Section~\ref{sec:method} describes the details of our proposed methods. Section~\ref{sec:experiments} discusses experimental results and ablation studies, while Section~\ref{sec:conclusion} gives a conclusion for this paper.

\section{Related Work}\label{sec:related_work}
\subsection{Traditional Video Coding}
Traditional video coding has been developed for several decades, leading to the creation of several widely adopted standards, such as H.264/AVC~\cite{wiegand2003overview}, H.265/HEVC~\cite{sullivan2012overview}, and H.266/VVC~\cite{bross2021overview}. These standards follow a similar hybrid coding framework, which typically integrates block-based motion compensated prediction~\cite{kim2012zoom,lin2013motion,chien2021motion}, transform~\cite{zhao2021transform,sole2012transform}, quantization~\cite{budagavi2014hevc,schwarz2021quantization,stankowski2015rate}, entropy coding~\cite{sze2012high,schwarz2021quantization}, and in-loop filtering~\cite{karczewicz2021vvc,norkin2012hevc}. While each successive generation has delivered substantial improvements in compression performance over its predecessor, the pace of improvement has gradually slowed, indicating a degree of maturity in this classical paradigm. In recent years, the remarkable success of deep learning has motivated the exploration of data-driven approaches, spurring significant interest and rapid progress in neural video coding as a promising complementary direction.
\begin{figure*}[t]
  \centering
   \includegraphics[width=0.86\linewidth]{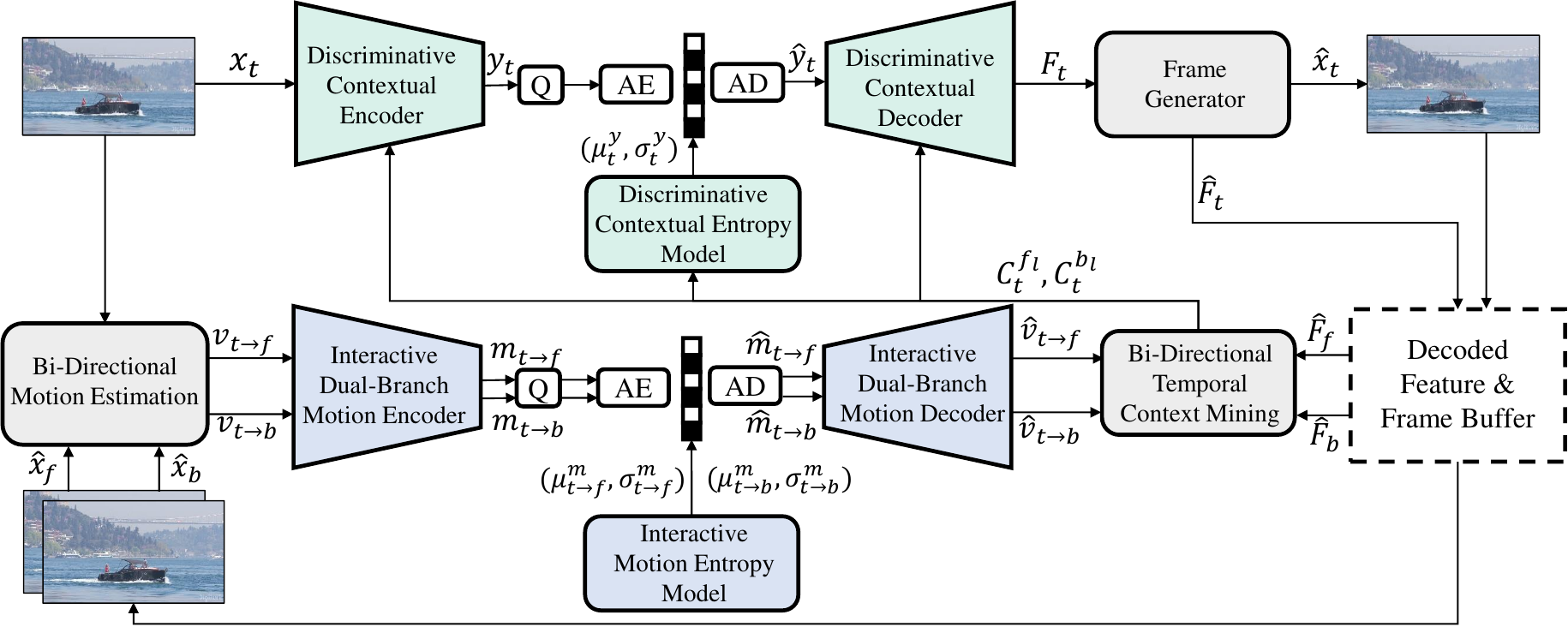}
      \caption{The framework of our proposed neural B-frame video codec, which consists of two major contributions: a fine-grained motion compression method with an interactive dual-branch auto-encoder and an interactive motion entropy model, and a selective temporal fusion method employing a discriminative contextual encoder-decoder and a discriminative contextual entropy model.}
   \label{fig:framework}
\end{figure*}
\subsection{Neural P-Frame Video Coding}
The majority of neural video codecs adopt the P-frame coding paradigm~\cite{lu2020end,sheng2024nvc,hu2022coarse,Rippel_2021_ICCV,hu2020improving,lu2020content,lin2020m,hu2021fvc,agustsson2020scale,cheng2019learning,rippel2019learned,liu2021deep,liu2020neural,liu2022end,yilmaz2021end,lin2022dmvc,guo2023learning,guo2023enhanced, liu2020conditional, sheng2025drfc,wang2023learned,wu2025end,wei2025rdvc,ma2024uncertainty,DBLP:conf/nips/MentzerTMCHLA22,li2021deep, ho2022canf,chen2024maskcrt,sheng2025prediction,sheng2024vnvc, sheng2022temporal,li2022hybrid,li2023neural,sheng2024spatial,li2024neural,chen2021nerv,wang2024ssnvc,kwan2024hinerv}.
Lu et al.~\cite{lu2020end} proposed a pioneering work in this field---DVC~\cite{lu2020end}, which implements key video coding components including motion estimation, motion compression, motion compensation, residual compression, and entropy models with neural networks and optimizes them in an end-to-end manner. Subsequently, Lin et al.~\cite{lin2020m} proposed to fuse the information of multiple previously compressed motion vectors to reduce the motion coding costs.  Agustsson et al.~\cite{agustsson2020scale} developed a scale-space flow, extending optical flow with a scale parameter to better model uncertainty, along with an enhanced warping operator robust to disocclusions and fast motion. Alternative to optical flow-based approaches, Hu et al.~\cite{hu2021fvc} proposed to represent motion with the offsets of deformable convolution and utilize deformable convolution to perform motion compensation. Based on this work, they further proposed to perform deformable convolution-based motion estimation and compensation in a coarse-to-fine manner~\cite{hu2022coarse}.  \par
The abovementioned schemes adopt a residual coding paradigm to utilize predicted temporal contexts. Beyond residual coding, Li et al.~\cite{li2021deep} proposed a conditional coding paradigm to better utilize temporal correlation, which fuses predicted temporal contexts into a contextual encoder-decoder and a contextual entropy model, enabling automatic learning of more effective methods for reducing temporal redundancy. Based on this scheme, Sheng et al.~\cite{sheng2022temporal} proposed to replace reference frames with reference features to propagate temporal information and designed a temporal context mining module that performs flow-based warping on these reference features to learn multi-scale temporal contexts. This method leverages spatio-temporal non-uniformity to generate more accurate temporal predictions.  Li et al.~\cite{li2022hybrid} further proposed a multi-granularity quantization mechanism to perform spatial-channel-wise quantization. This content-adaptive quantization mechanism not only helps achieve the smooth rate adjustment in a single model but also improves the compression performance by dynamic bit allocation. Through integration of these advanced techniques, recent DCVC-based learned P-frame codecs~\cite{li2024neural,li2023neural,sheng2025prediction} have outperformed the reference software of traditional video coding standard H.266/VVC~\cite{bross2021overview} under low-delay configurations by a large margin.\par

\subsection{Neural B-Frame Video Coding}
The development of neural B-frame video coding~\cite{wu2018video,nguyen2024motion,nguyenquang2025fast, alexandre2023hierarchical, cheng2019learning, xu2024ibvc,gao2025maskcrt, yang2020learning,yang2021learning,yang2022advancing,yilmaz2020end,chen2023b,feng2021versatile,yang2024ucvc,djelouah2019neural,pourreza2021extending,sheng2025bi,lin2024hierarchical,kim2023neural}  is relatively slower compared with neural P-frame video coding. Based on the need for motion coding, existing neural B-frame coding schemes can be categorized into two types.\par
Schemes in the first category~\cite{wu2018video,nguyen2024motion, alexandre2023hierarchical, cheng2019learning, xu2024ibvc} depend on video interpolation to perform bi-directional temporal prediction without the need for motion coding. Wu~\emph{et al.}~\cite{wu2018video} designed a contextual video interpolation network that uses bi-directional multi-scale contexts to obtain predicted frames.  Nguyen \emph{et al.}~\cite{alexandre2023hierarchical} adopted a kernel-based, motion-free approach for neural video compression, where video predictions are generated from reference frames via kernel-based interpolation. Cheng \emph{et al.}~\cite{cheng2019learning} proposed to adaptively select the frame number in one interpolation loop according to video motion patterns. Xu~\cite{xu2024ibvc} proposed to select interpolated temporal contexts adaptively using a residual-guided masking encoder. Alexandre \emph{et al.}~\cite{alexandre2023hierarchical} introduced a two-layer neural B-frame codec that merges the low-resolution interpolated frames from the base layer and high-resolution interpolated frames from the enhancement layer to generate more accurate predicted frames. Currently, the best interpolation-based neural B-frame video codec~\cite{alexandre2023hierarchical} has outperformed x265---an efficient and widely used H.265/HEVC encoding software implementation. \par

Schemes in the second category~\cite{yang2020learning,gao2025maskcrt,nguyenquang2025fast,yang2021learning,yang2022advancing,yilmaz2020end,chen2023b,feng2021versatile,yang2024ucvc,djelouah2019neural,pourreza2021extending,sheng2025bi,lin2024hierarchical,kim2023neural} aim to enhance bi-directional temporal prediction by compressing bi-directional motion information. For example, Yang \emph{et al.}~\cite{yang2020learning} proposed to perform bi-directional motion vector estimation and compression to obtain bi-directional predicted frames.  Subsequently, they employed a residual codec to compress the prediction residuals. Feng \emph{et al.}~\cite{feng2021versatile} proposed to replace the bi-directional optical flows with a voxel flow and replace the bi-directional prediction merging process with the trilinear warping operation. Recently, some schemes~\cite{yang2024ucvc,chen2023b,gao2025maskcrt,lin2024hierarchical,sheng2025bi,kim2023neural} proposed to extend conditional coding~\cite{li2021deep} from P-frame to B-frame. For instance, Gao \emph{et al.}~\cite{gao2025maskcrt} introduced MaskCRT-B, a B-frame coding method based on a masked conditional residual transformer. It employs conditional residual coding and a bi-directional adaptive fusion module to enhance the prediction of occluded and disoccluded regions. Building upon this work, NguyenQuang \emph{et al.}~\cite{nguyenquang2025fast} developed a fast online motion resolution adaptation method. This method employs lightweight classifiers to determine the optimal motion downsampling factor, thereby addressing failures in large motion prediction.
Recently, Sheng \emph{et al.}~\cite{sheng2025bi} proposed DCVC-B that predicts bi-directional multi-scale temporal contexts using a bi-directional temporal context mining module. These contexts are then directly fused into a contextual encoder-decoder and a contextual entropy model to reduce temporal redundancy.
Although DCVC-B~\cite{sheng2025bi} surpassed the reference software of H.265/HEVC standard with random-access configurations, it is difficult to achieve higher B-frame compression performance by simply applying P-frame coding methods to B-frame coding without fully considering the unique characteristics of bi-directional prediction.

\section{Methodology}\label{sec:method}
We begin with a framework overview, highlighting the relationship to the baseline DCVC-B~\cite{sheng2025bi}, and then detail our proposed fine-grained motion compression and selective temporal fusion methods. 
\subsection{Framework Overview}
The architecture of our neural B-frame video codec is illustrated in Fig.~\ref{fig:framework}. While following a similar architecture to DCVC-B~\cite{sheng2025bi}, our work introduces key innovations to address specific limitations in B-frame coding. The framework retains several components from the baseline, including bi-directional motion estimation, temporal context mining, and frame generation. Our main contributions focus on substantially improving the motion compression and contextual compression.
First, to reduce bi-directional motion coding costs, we propose a fine-grained motion compression method that replaces the original coarse-grained motion compression with an interactive dual-branch auto-encoder, per-branch adaptive quantization, and an interactive motion entropy model. Second, to enable discriminative utilization of temporal contexts and priors in the contextual compression, we develop a selective temporal fusion method that supersedes the original uniform fusion strategy through bi-directional weighting and implicit prior alignment. These key innovations are designed to tackle the unique challenges of B-frame coding, leading to substantial gains in compression efficiency.
\begin{figure}[t]
  \centering
   \includegraphics[width=\linewidth]{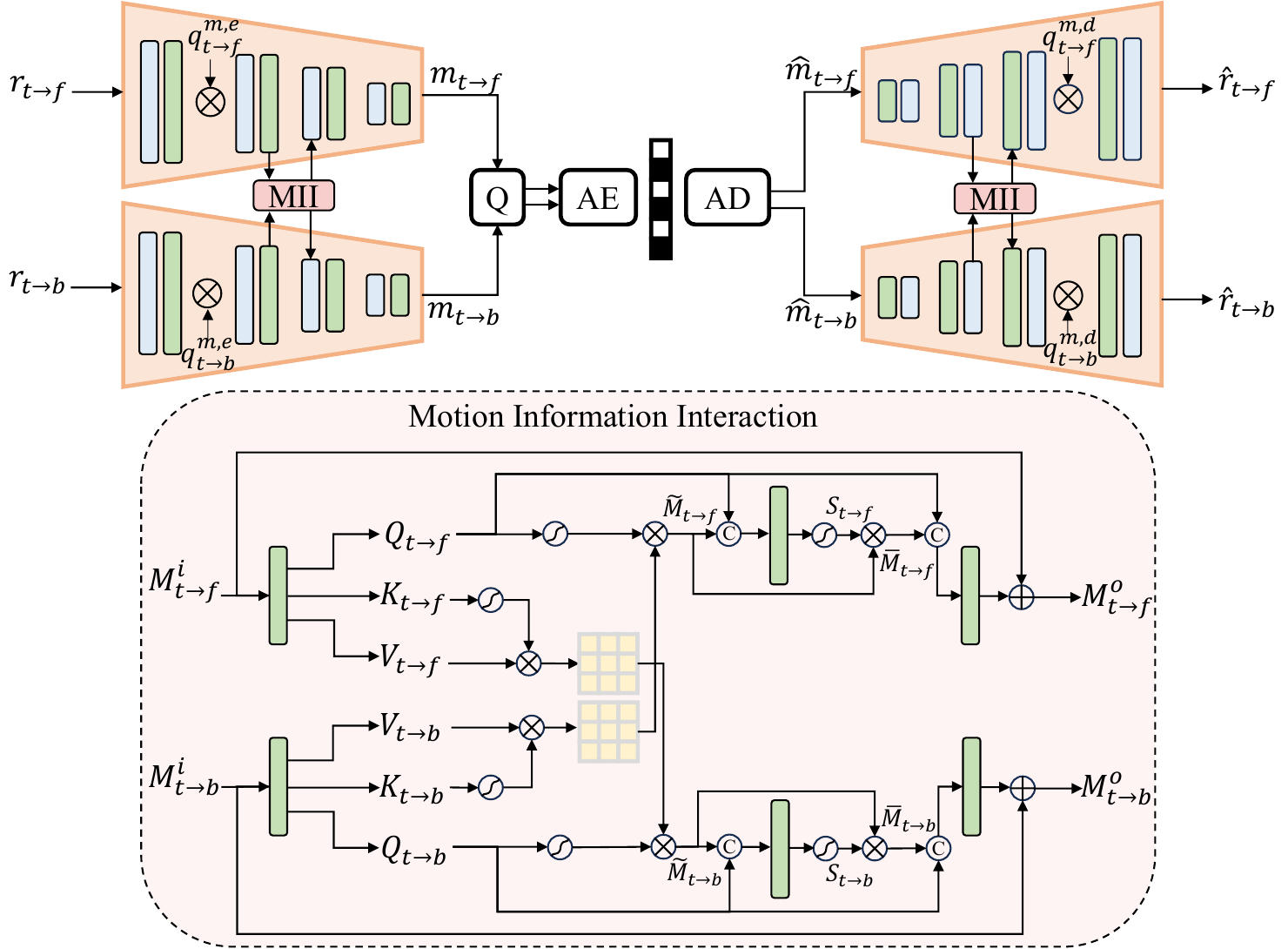}
      \caption{Illustration of our proposed interactive dual-branch motion auto-encoder with per-branch adaptive quantization steps. ``Q" refers to the rounding-based quantization operator. ``AE" and ``AD" refer to the arithmetic encoder and decoder, respectively. $q_{t\rightarrow f}^{m,e}$, $q_{t\rightarrow b}^{m,e}$, $q_{t\rightarrow f}^{m,d}$, and $q_{t\rightarrow b}^{m,d}$ are learnable quantization steps.}
   \label{fig:DualBranchMED}
\end{figure}
\subsection{Fine-Grained Motion Compression}
We employ SpyNet~\cite{ranjan2017optical} to estimate bi-directional motion vectors ($v_{t\rightarrow f}$, $v_{t\rightarrow b}$) and associated predictions ($\frac{v_{b\rightarrow f}}{2}$, $\frac{v_{f\rightarrow b}}{2}$). Our fine-grained motion compression method is used to encode and decode their motion vector differences ($r_{t\rightarrow f}$, $r_{t\rightarrow b}$)~\cite{sheng2025bi}. 

\subsubsection{Interactive Dual-Branch Motion Auto-Encoder}
Our method incorporates an interactive dual-branch motion auto-encoder, illustrated in Fig.~\ref{fig:DualBranchMED}. Unlike the scheme that concatenated bi-directional motion vector differences (MVDs) for joint compression~\cite{sheng2025bi}, our design processes forward and backward MVDs 
($r_{t\rightarrow f}$, $r_{t\rightarrow b}$) separately. This is motivated by the observation that bi-directional reference frames often exhibit asymmetric temporal correlations, necessitating distinct motion reconstruction qualities for each direction.
The dual-branch motion encoder---comprising downsampling residual blocks and depth-wise convolution blocks---compresses MVDs into two latent representations ($m_{t\rightarrow f}$, $m_{t\rightarrow b}$).  These representations are quantized and entropy encoded into bitstreams. During decoding, the dual-branch motion decoder---comprising upsampling residual blocks and depth-wise convolution blocks---reconstructs the quantized latent representations ($\hat{m}_{t\rightarrow f}$, $\hat{m}_{t\rightarrow b}$) into motion vectors ($\hat{v}_{t\rightarrow f}$, $\hat{v}_{t\rightarrow b}$).\par

Unlike the scheme that employed a parameter-shared motion auto-encoder for separate bi-directional motion compression~\cite{ye2024deep}, our dual-branch motion auto-encoder designs a motion information interaction (MII) module to exploit the inherent geometric consistency between forward and backward motion vectors. As shown in Fig.~\ref{fig:DualBranchMED}, the motion information interaction module takes intermediate bi-directional motion features ($M_{t\rightarrow f}^{i}$, $M_{t\rightarrow b}^{i}$) as inputs and outputs ($M_{t\rightarrow f}^{o}$, $M_{t\rightarrow b}^{o}$). At the start of the MII module, two depth-wise convolution blocks $\left(k_{f}\left(\cdot\right), k_{b}\left(\cdot\right)\right)$ are separately applied to ($M_{t\rightarrow f}^{i}$, $M_{t\rightarrow b}^{i}$) to generate bi-directional query (Q), key (K), and value (V) features ($Q_{t\rightarrow f}, K_{t\rightarrow f}, V_{t\rightarrow f}$),  ($Q_{t\rightarrow b}, K_{t\rightarrow b}, V_{t\rightarrow b}$). 
\begin{equation}
\begin{aligned}
Q_{t\rightarrow f}, K_{t\rightarrow f}, V_{t\rightarrow f} &= k_{f}\left(M_{t\rightarrow f}^{i}\right),\\
Q_{t\rightarrow b}, K_{t\rightarrow b}, V_{t\rightarrow b} &= k_{b}\left(M_{t\rightarrow b}^{i}\right).
\end{aligned}
\end{equation}
Then we use an efficient attention mechanism~\cite{shen2021efficient} to exchange the bi-directional motion information, obtaining the initial cross-directional motion contexts ($\tilde{M}_{t \rightarrow f}, \tilde{M}_{t \rightarrow b}$).
\begin{equation}
\begin{aligned}
\tilde{M}_{t \rightarrow f}=\left(\sigma\left({K}_{t \rightarrow b}\right) \times {V}_{t \rightarrow b}^T\right)^T \times \sigma\left({Q}_{t \rightarrow f}\right),\\
\tilde{M}_{t \rightarrow b}=\left(\sigma\left({K}_{t \rightarrow f}\right) \times {V}_{t \rightarrow f}^T\right)^T \times \sigma\left({Q}_{t \rightarrow b}\right),
\end{aligned}
\end{equation}
where $\sigma$ is the softmax function. Subsequently, we obtain the refined motion contexts $(\bar{M}_{t \rightarrow f}, \bar{M}_{t \rightarrow b})$ by screening ($\tilde{M}_{t \rightarrow f}, \tilde{M}_{t \rightarrow b}$) according to ($Q_{t\rightarrow f}$, $Q_{t\rightarrow b}$).
\begin{equation}
\begin{aligned}
&\bar{M}_{t \rightarrow f}=\tilde{M}_{t \rightarrow f} \otimes  S_{t \rightarrow f},\\
&\bar{M}_{t \rightarrow b}=\tilde{M}_{t \rightarrow b} \otimes  S_{t \rightarrow b},\\
& \quad \text { with } \quad S_{t \rightarrow f} = \sigma\left(h_{f}\left(\tilde{M}_{t \rightarrow f}\textcircled{c}{Q}_{t \rightarrow f}\right)\right),\\
& \quad \text { with } \quad S_{t \rightarrow b} = \sigma\left(h_{b}\left(\tilde{M}_{t \rightarrow b}\textcircled{c}{Q}_{t \rightarrow b}\right)\right),
\end{aligned}
\end{equation}
where $h_{f}(\cdot)$ and $h_{b}(\cdot)$ are two depth-wise convolution blocks. Finally, $M_{t\rightarrow f}^{i}$ and $M_{t\rightarrow b}^{i}$ are transformed
 to more compact motion feature $M_{t\rightarrow f}^{o}$ and $M_{t\rightarrow b}^{o}$ conditioned on the cross-directional motion contexts  $(\bar{M}_{t \rightarrow f}, \bar{M}_{t \rightarrow b})$.
\begin{equation}
\begin{aligned}
{M}_{t \rightarrow f}^{o}&=M_{t\rightarrow f}^{i}+\left(g_{f}\left(\bar{M}_{t \rightarrow f}\textcircled{c}{Q}_{t \rightarrow f}\right)\right),\\
{M}_{t \rightarrow b}^{o}&=M_{t\rightarrow b}^{i}+\left(g_{b}\left(\bar{M}_{t \rightarrow b}\textcircled{c}{Q}_{t \rightarrow b}\right)\right),
\end{aligned}
\end{equation}
where $g_{f}(\cdot)$ and $g_{b}(\cdot)$ are two depth-wise convolution blocks. $\textcircled{c}$ refers to channel-wise concatenation. With our proposed MII module, the information from dual motion branches can be exchanged, thereby exploiting the cross-directional motion dependencies and reducing the motion coding costs.

\subsubsection{Per-Branch Adaptive Quantization Steps}
Adaptive quantization step techniques~\cite{li2022hybrid, li2023neural,li2024neural} have proven effective for precise rate-distortion control in neural video coding. While recent work~\cite{sheng2025bi} successfully extended this method from P-frame to B-frame coding, it employed identical quantization steps for bi-directional motion vectors. This uniform treatment fails to account for the distinct bitrate allocation needs and reconstruction quality requirements inherent to bi-directional motion vectors.
To address this limitation, we propose to integrate independent learnable quantization steps into each branch of both the motion encoder and decoder, as presented in Fig.~\ref{fig:DualBranchMED}. One quantization step is composed of a global quantization step and a channel-wise quantization step. The global quantization step is a number related to the target bit rate. The channel-wise quantization step is a one-dimensional vector, and the length of the vector is the same as the number of channels of the quantized feature.
\begin{equation}
\begin{aligned}
q_{t\rightarrow f}^{m,e} &= q_{t\rightarrow f}^{m,e,global}\odot q_{t\rightarrow f}^{m,e,ch},\\
q_{t\rightarrow f}^{m,d} &= q_{t\rightarrow f}^{m,d,global}\odot q_{t\rightarrow f}^{m,d,ch},\\
q_{t\rightarrow b}^{m,e} &= q_{t\rightarrow b}^{m,e,global}\odot q_{t\rightarrow b}^{m,e,ch},\\
q_{t\rightarrow b}^{m,d} &= q_{t\rightarrow b}^{m,d,global}\odot q_{t\rightarrow b}^{m,d,ch},\\
\end{aligned}
\end{equation}
where $\odot$ refers to element-wise multiplication. This design enables fine-grained, direction-specific rate-distortion control for forward and backward motion vectors, optimizing compression efficiency for B-frame coding.
\begin{figure}[t]
  \centering
   \includegraphics[width=0.6\linewidth]{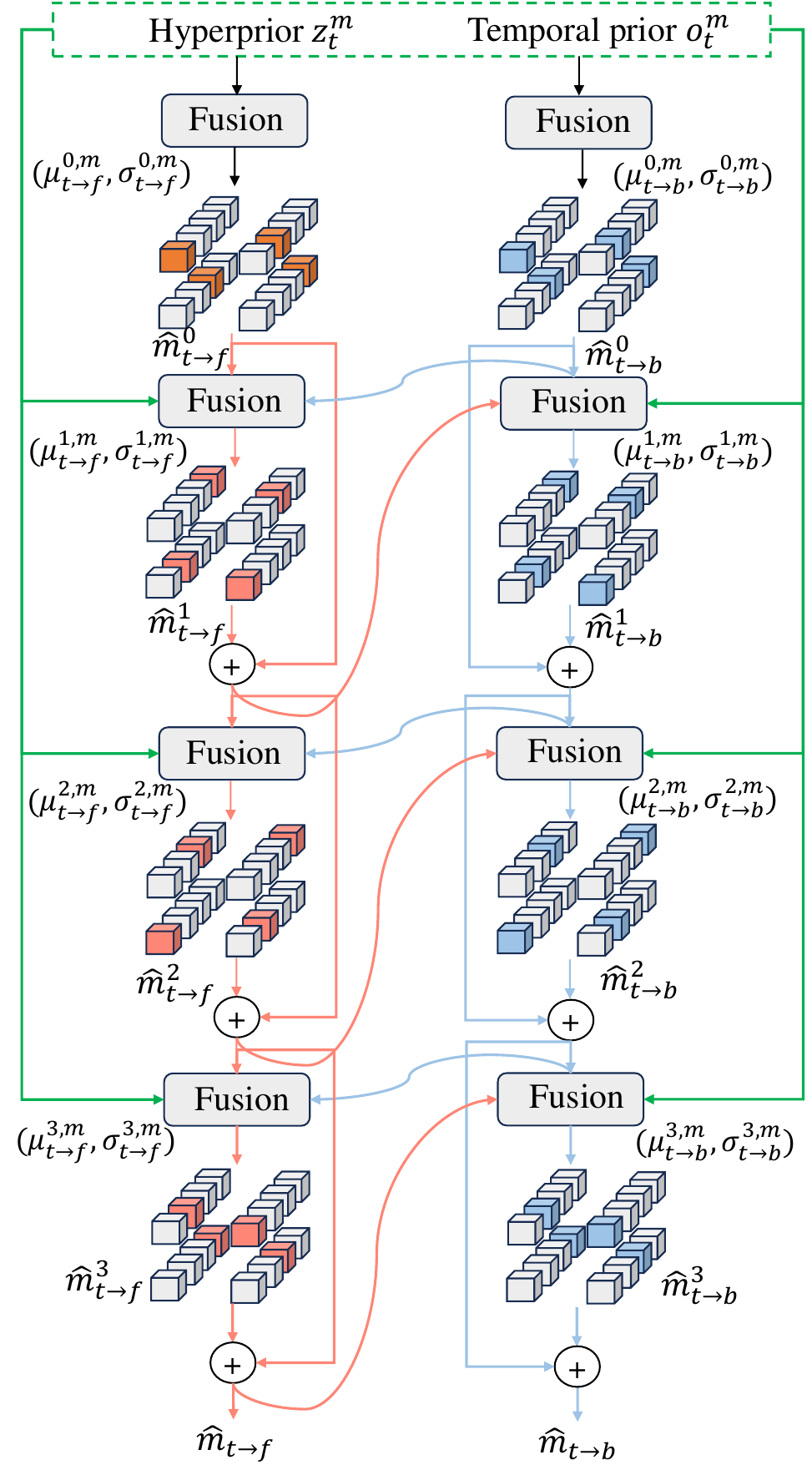}
      \caption{Illustration of our proposed interactive motion entropy model. }
   \label{fig:entropy_model}
\end{figure}
\subsubsection{Interactive Motion Entropy Model}
Given the quantized bi-directional motion latent representations ($\hat{m}_{t\rightarrow f}$, $\hat{m}_{t\rightarrow b}$), we propose an interactive motion entropy model to estimate their probability distributions for arithmetic coding. First, we derive a motion hyperprior $z_t^m$ through a motion hyper encoder $\mathcal{E}_h^m$ and a motion hyper decoder $\mathcal{D}_h^m$:
\begin{equation}
z_t^m = \mathcal{D}_h^m\left(\mathcal{E}_h^m \left(\hat{m}_{t\rightarrow f}\textcircled{c}\hat{m}_{t\rightarrow b}\right)\right).
\end{equation}
Then, we use a motion temporal prior extractor $\mathcal{T}^m$~\cite{sheng2025bi} to leverage the temporal correlation in motion vectors. 
\begin{equation}
o_t^m = \mathcal{T}^m \left(v_{b\rightarrow f}\textcircled{c}v_{f\rightarrow b}\right).
\end{equation}
The extractor $\mathcal{T}^m$, composed of four convolutional layers with downsampling, takes the concatenated bi-directional motion vectors ($v_{b\rightarrow f}$, $v_{f\rightarrow b}$) as inputs and generates a motion temporal prior $o_t^m$. The output $o_t^m$ has the same spatial resolution and channel number as the motion latent representations ($\hat{m}_{t\rightarrow f}$, $\hat{m}_{t\rightarrow b}$), allowing it to serve as a direct motion temporal prior in the motion entropy model.

\par

Next, we partition $\hat{m}_{t\rightarrow f}$ and $\hat{m}_{t\rightarrow b}$ into four segments via spatial-channel splitting~\cite{li2022hybrid, li2023neural,li2024neural,sheng2025bi,sheng2025prediction,sheng2024spatial}, as illustrated in Fig.~\ref{fig:entropy_model}.
To effectively capture bi-directional motion dependencies, when compressing the current motion latent segment $\hat{m}_{t\rightarrow f}^{i}$ ($\hat{m}_{t\rightarrow b}^{i}$), we not only fuse the motion hyperprior $z_t^m$, motion temporal prior $o_t^m$, and previously compressed latent segments $\hat{m}_{t\rightarrow f}^{<i}$ ($\hat{m}_{t\rightarrow b}^{<i}$) in the same direction, but also fuse the previously compressed latent segments in the other direction $\hat{m}_{t\rightarrow b}^{<i}$ ($\hat{m}_{t\rightarrow f}^{<i+1}$).
\begin{equation}
\begin{aligned}
& p_{\hat{m}_{t\rightarrow f}^{i}}\left(\hat{m}_{t\rightarrow f}^{i} \mid  z_t^m, o_t^m, \hat{m}_{t\rightarrow f}^{<i}, \hat{m}_{t\rightarrow b}^{<i} \right) \sim \mathcal{L}\left(\mu_{t\rightarrow f}^{m,i}, \sigma_{t\rightarrow f}^{m,i}\right), \\
& p_{\hat{m}_{t\rightarrow b}^{i}}\left(\hat{m}_{t\rightarrow b}^{i} \mid  z_t^m, o_t^m, \hat{m}_{t\rightarrow b}^{<i}, \hat{m}_{t\rightarrow f}^{<i+1} \right) \sim \mathcal{L}\left(\mu_{t\rightarrow b}^{m,i}, \sigma_{t\rightarrow b}^{m,i}\right), \\
\end{aligned}
\end{equation}

where $\mathcal{L}\left(\mu, \sigma\right)$ denotes a Laplacian distribution. Our entropy model achieves fine-grained motion dependency modeling by interactively fusing bi-directional latent segments as priors. This strategy, while introducing a partial sequential dependency that limits full parallel decoding across the two directions of motion streams, effectively exploits the inherent geometric correlations in bi-directional motion vectors. By leveraging cross-directional motion prior information, it achieves more accurate motion entropy estimation and enhanced compression performance.

\begin{figure}[t]
  \centering
   \includegraphics[width=\linewidth]{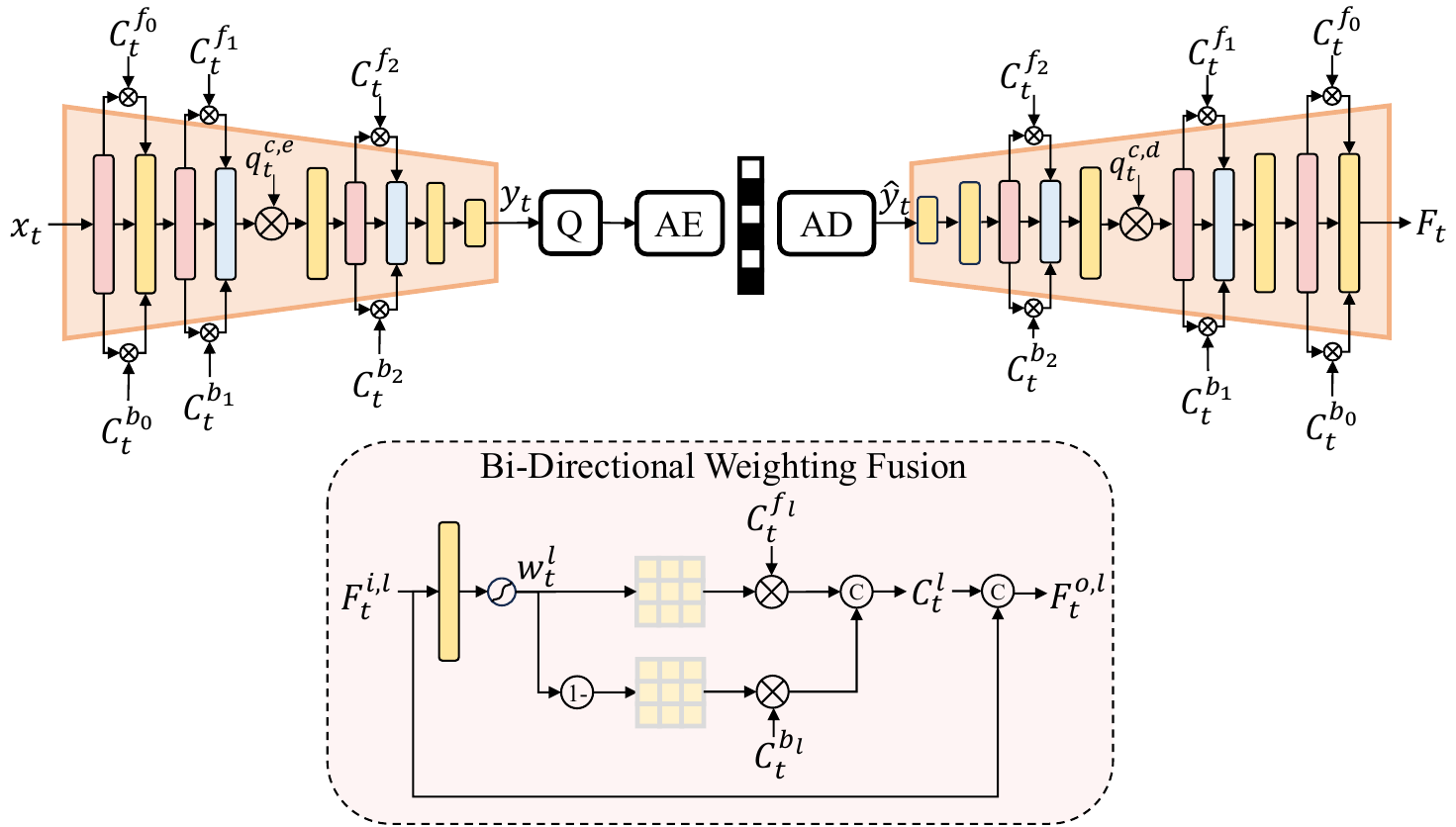}
      \caption{Illustration of our proposed contextual auto-encoder with bi-directional weighting-based context fusion. ``Q" refers to the rounding-based quantization operator. ``AE" and ``AD" refer to the arithmetic encoder and decoder, respectively. $q_{t}^{c,e}$ and $q_{t}^{c,d}$ are learnable quantization steps.}
   \label{fig:CED}
\end{figure}
\subsection{Selective Temporal Fusion}
Given the reconstructed bi-directional motion vectors, we employ the temporal context mining moulde~\cite{sheng2022temporal} to extract multi-scale temporal contexts from reference features ($\hat{F}_f, \hat{F}_b$), yielding forward contexts $(C_t^{f_0},C_t^{f_1},C_t^{f_2})$ and and backward contexts $(C_t^{b_0},C_t^{b_1},C_t^{b_2})$, as illustrated in Fig.~\ref{fig:framework}. These contexts and previously compressed contextual latent representations ($\hat{y}_f, \hat{y}_b$) are fused into the contextual encoder-decoder and context entropy model to reduce temporal redundancy. Our selective temporal fusion method focuses on how to discriminatively utilize these bi-directional temporal contexts and temporal priors.

\subsubsection{Bi-Directional Weighting-based Context Fusion}
Our method first incorporates a bi-directional weighting fusion mechanism to build a discriminative contextual encoder-decoder. As illustrated in Fig.~\ref{fig:CED}, unlike the schemes~\cite{sheng2025bi} that blindly fused bi-directional temporal contexts, our discriminative contextual encoder-decoder predicts bi-directional fusion weights to  selectively fuse the intermediate contextual feature $F_t^{i,l}$ with the bi-directional temporal contexts ($C_t^{f_l}, C_t^{b_l}$).
\begin{equation}
\begin{aligned}
&F_t^{o,l}=F_t^{i,l}\textcircled{c}\left({w}_{t}^{f_l} \odot C_t^{f_l}\right)\textcircled{c}\left({w}_{t}^{b_l} \odot C_t^{b_l}\right),\\
& \quad \text { with } \quad {w}_{t}^{f_l}=\sigma\left(f_l\left(F_t^l\right)\right),\\
& \quad \text { with } \quad {w}_{t}^{f_b}=1-{w}_{t}^{f_l}, \\
\end{aligned}
\end{equation}
where $f_l$ is a convolutional layer. Through the end-to-end optimization, the predicted fusion weights adaptively recalibrate the contributions of bi-directional temporal contexts based on their relevance to the target frame, thereby enabling selective temporal fusion that suppresses redundant or noisy temporal contexts.
\begin{figure}[t]
  \centering
   \includegraphics[width=0.9\linewidth]{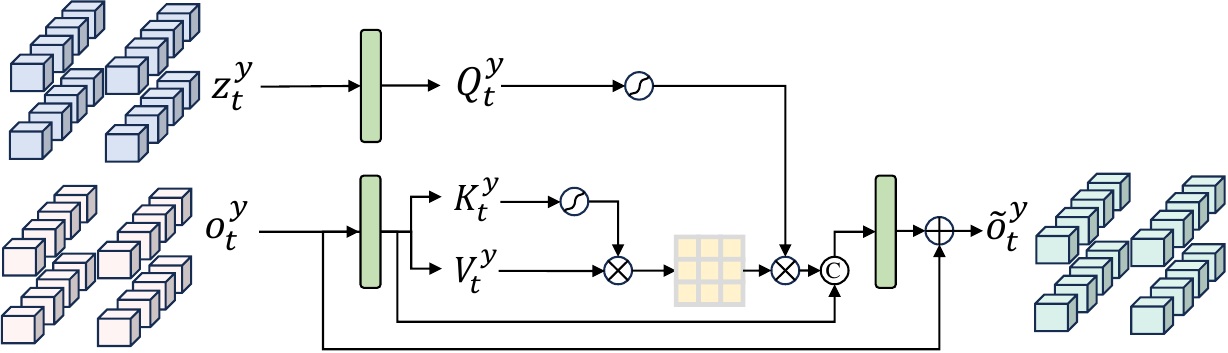}
      \caption{Illustration of our proposed implicit alignment-based prior fusion.}
   \label{fig:ContextualEntropyModel}
\end{figure}
\subsubsection{Implicit Alignment-based Prior Fusion}
Previous schemes~\cite{sheng2025bi} tended to fuse the compressed contextual latent representations ($\hat{y}_{f}$, $\hat{y}_{b}$) and small resolution of temporal contexts ($C_t^{f,2}$, $C_t^{b,2}$) into the contextual entropy model as a temporal prior $o_t^{y}$. However, these schemes ignore the spatial misalignment of latent representations and the predicted errors of temporal contexts, which fail to optimally exploit temporal correlations. Therefore, our method further incorporates a simple hyperprior-based implicit alignment for the contextual entropy model. Specifically, we first use two depth-wise convolution blocks $\left(s_{z}\left(\cdot\right), s_{o}\left(\cdot\right)\right)$ to generate query, key, and value features ($Q_{t}^y, K_{t}^y, V_{t}^y$) from hyperprior $z_t^{y}$ and temporal prior  $o_t^{y}$.
\begin{equation}
\begin{aligned}
Q_{t}^y &= s_{z}\left(z_t^{y}\right),\\
K_{t}^y, V_{t}^y &= s_{o}\left(o_t^{y}\right),\\
\end{aligned}
\end{equation}
Considering the hyperprior $z_t^{y}$ is an estimation of the current contextual latent representation $\hat{y}_t$, we treat it as a surrogate to help align the temporal prior $o_t^{y}$ using an efficient attention mechanism~\cite{shen2021efficient}. 
\begin{equation}
\begin{aligned}
\bar{o}_{t}^y=\left(\sigma\left(K_{t}^y\right) \times \left({V}_{t}^y\right)^T\right)^T \times \sigma\left({Q}_{t}^y\right).\\
\end{aligned}
\end{equation}
Then, we output the refined temporal prior $\tilde{o}_{t}^{y}$ using the aligned feature.
\begin{equation}
\tilde{o}_{t}^{y}=o_{t}^{y}+\left(j_{o}\left(\bar{o}_{t}^y\textcircled{c}s_{o}\left(o_t^{y}\right)\right)\right),\\
\end{equation}
where $j_o$ is a depth-wise convolution block. The aligned temporal prior is then fused with the hyperprior and spatial prior to achieve discriminative entropy modeling.
\begin{figure*}[t]
  \begin{minipage}[c]{0.31\linewidth}
  \centering
    \includegraphics[width=\linewidth]{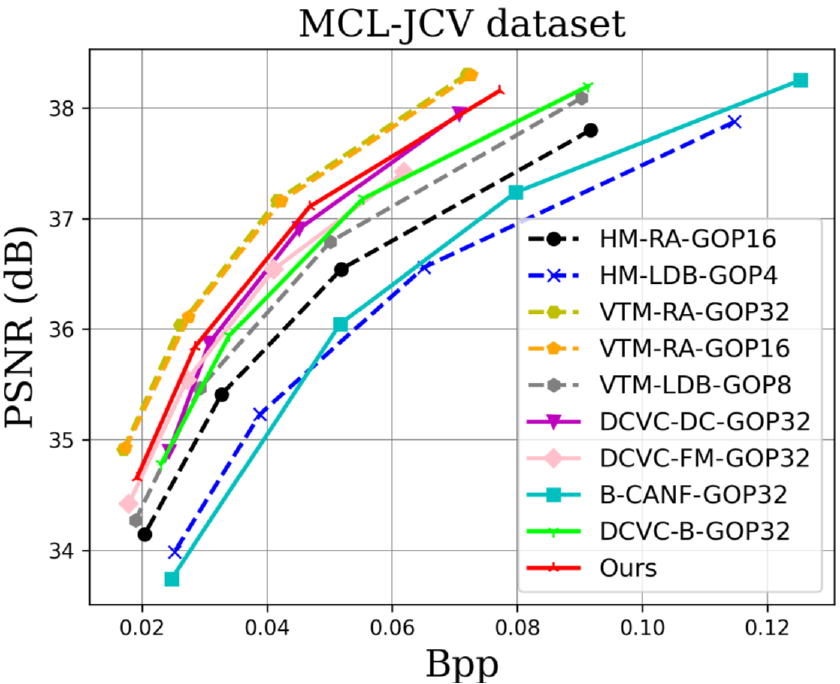}
  \end{minipage}%
  \begin{minipage}[c]{0.31\linewidth}
  \centering
    \includegraphics[width=\linewidth]{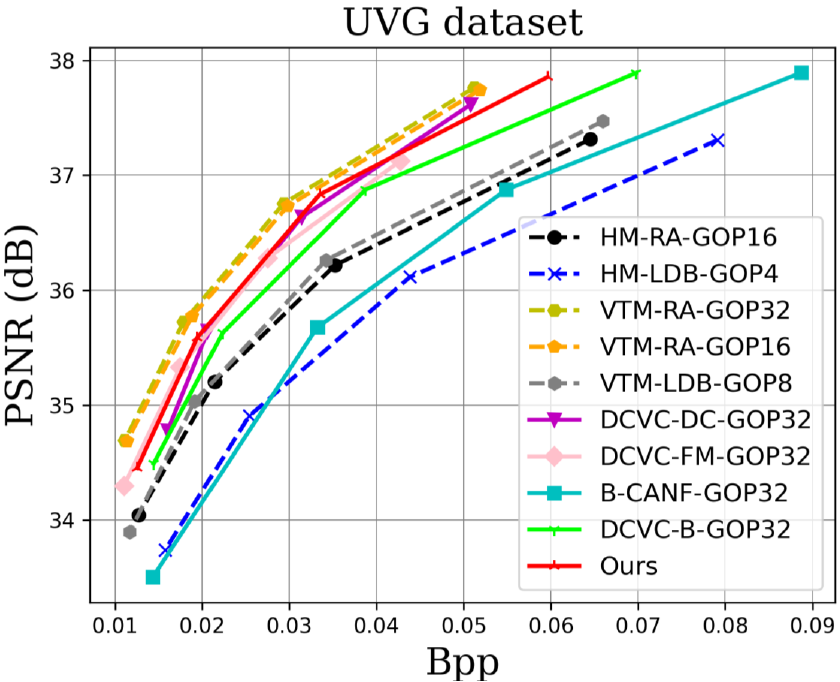}
  \end{minipage}%
  \centering
  \begin{minipage}[c]{0.31\linewidth}
  \centering
  \includegraphics[width=\linewidth]{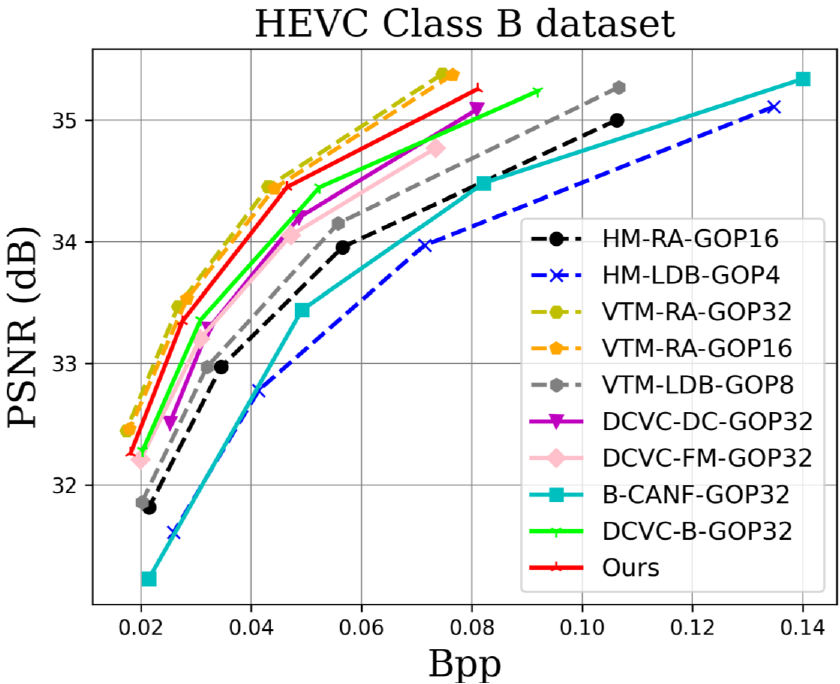}
 \end{minipage}%

  \begin{minipage}[c]{0.31\linewidth}
  \centering
    \includegraphics[width=\linewidth]{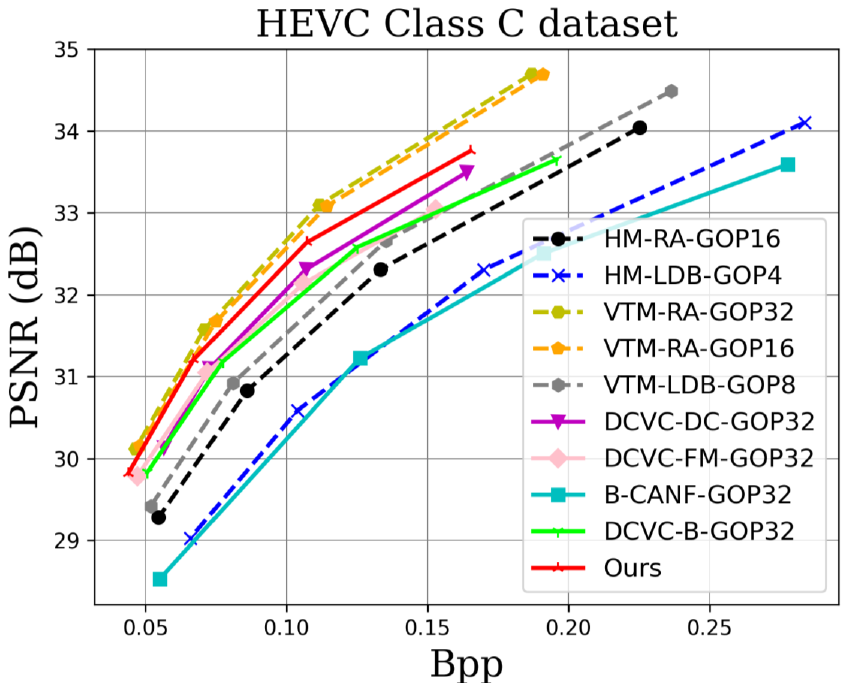}
  \end{minipage}%
  \begin{minipage}[c]{0.31\linewidth}
  \centering
    \includegraphics[width=\linewidth]{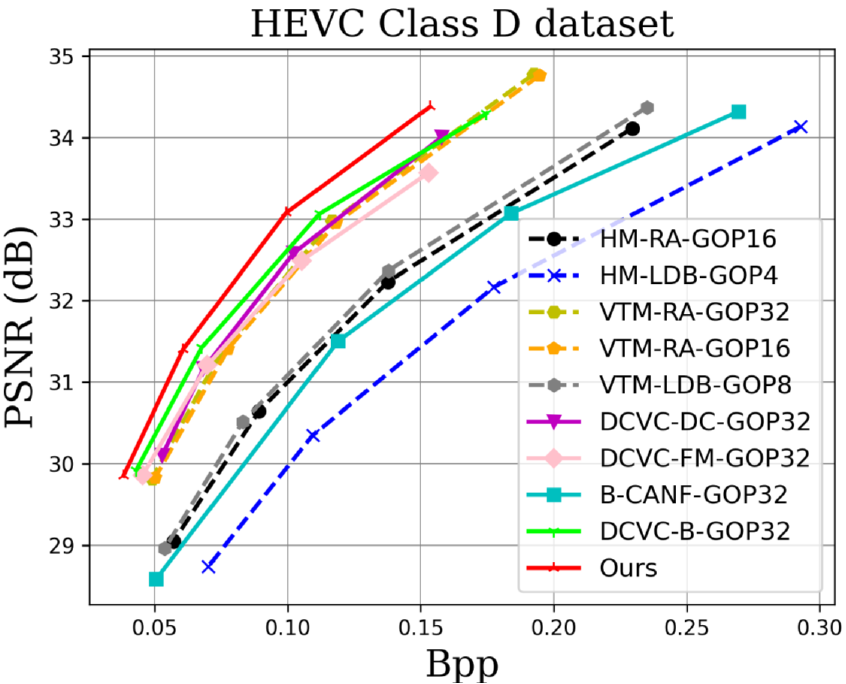}
  \end{minipage}%
  \begin{minipage}[c]{0.31\linewidth}
  \centering
    \includegraphics[width=\linewidth]{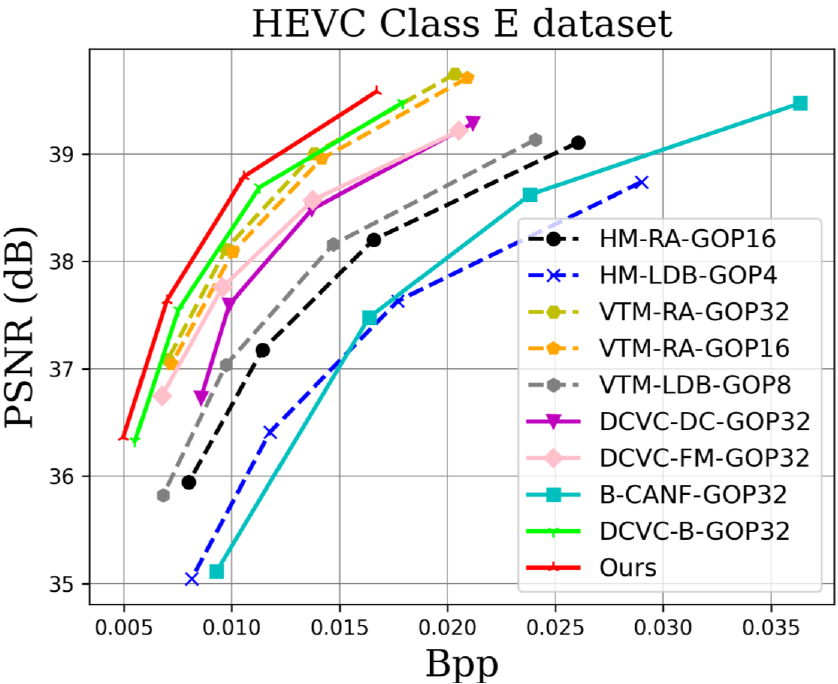}
  \end{minipage}%

    \caption{RD curves of various codecs on MCL-JCV, UVG, and HEVC datasets, measured by PSNR over 96 testing frames.}
  \label{fig:psnr_results}
\end{figure*}

\begin{table*}[t]
\caption{Comparison of BD-rate (\%) against HM-RA-GOP16 using PSNR for reconstruction quality over 96 testing frames. } 
  \centering
\scalebox{0.9}{
\begin{threeparttable}
\begin{tabular}{l|c|c|c|c|c|c|c}
\toprule[1.5pt]
                        &MCL-JCV &UVG & HEVC Class B  & HEVC Class C  &HEVC Class D &HEVC Class E&Average\\ \hline
HM-RA-GOP16             &0.0     &0.0     &0.0          &0.0          &0.0          &0.0      &0.0   \\ \hline
HM-LDB-GOP4             &23.9    &29.8    &23.7         &26.2         &29.7         &32.2     &27.6   \\ \hline
VTM-RA-GOP32            &\textbf{--38.0}  &\textbf{--35.9}  &\textbf{--39.1}       &\textbf{--33.7}       &--30.4       &--37.9   &\textbf{--35.8}   \\ \hline
VTM-RA-GOP16            &--36.9  &--34.0  &--37.2       &--32.1       &--29.2       &--35.4   &--34.1   \\ \hline
VTM-LDB-GOP8            &--13.3  &--5.3   &--11.7       &--8.6        &--3.9        &--10.4   &--8.9   \\ \hline
DCVC-DC-GOP32           &--22.8  &--25.0  &--22.0       &--19.9       &--32.3       &--24.6   &--24.4   \\ \hline
DCVC-FM-GOP32           &--20.7  &--23.7  &--20.5       &--19.3       &--31.1       &--30.5   &--24.3   \\ \hline
B-CANF-GOP32            &22.6    &21.0    &13.9         &30.9         &5.1          &28.5     &20.3   \\ \hline
DCVC-B-GOP32            &--16.8  &--18.0  &--26.8       &--15.8       &--36.7       &--42.9   &--26.1   \\ \hline
Ours-GOP32              &--27.5  &--27.4  &--34.7       &--28.4       &\textbf{--43.5}       &\textbf{--49.0}   &--35.1   \\ 
\bottomrule[1.5pt]
\end{tabular}
\end{threeparttable}
}
\label{table:ip32_psnr}
\end{table*}
\begin{figure*}[t]
  \begin{minipage}[c]{0.31\linewidth}
  \centering
    \includegraphics[width=\linewidth]{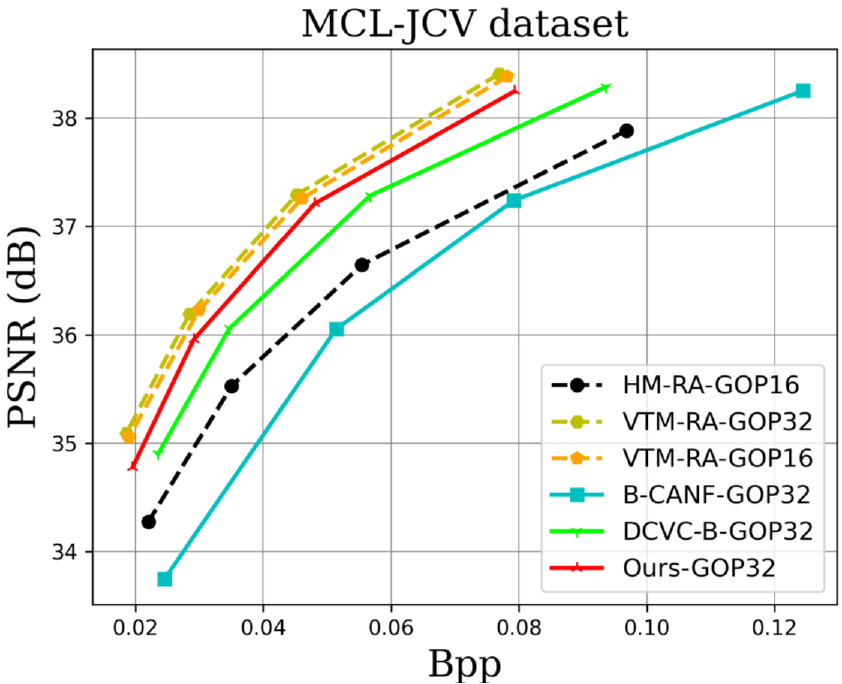}
  \end{minipage}%
  \begin{minipage}[c]{0.31\linewidth}
  \centering
    \includegraphics[width=\linewidth]{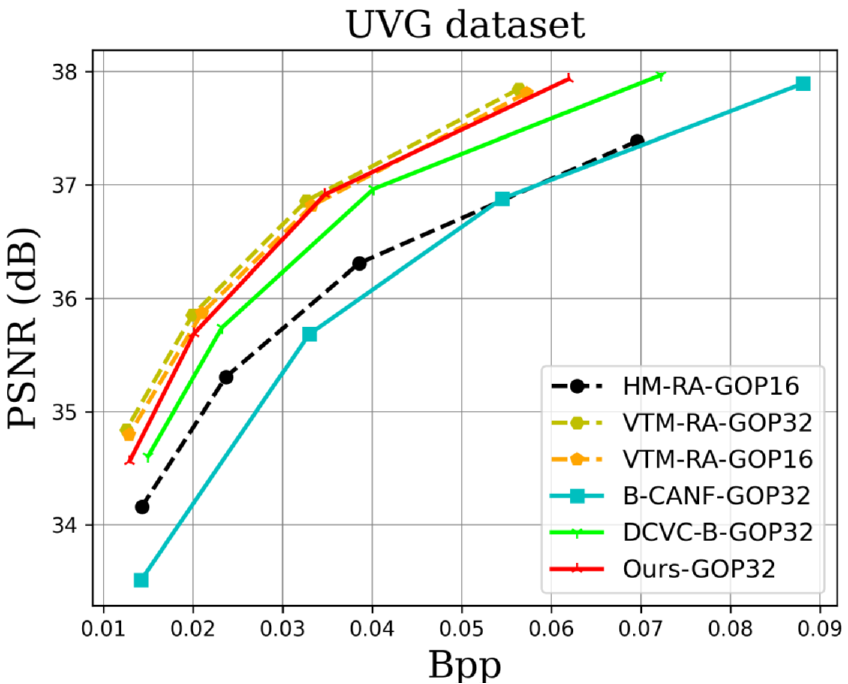}
  \end{minipage}%
  \centering
  \begin{minipage}[c]{0.31\linewidth}
  \centering
  \includegraphics[width=\linewidth]{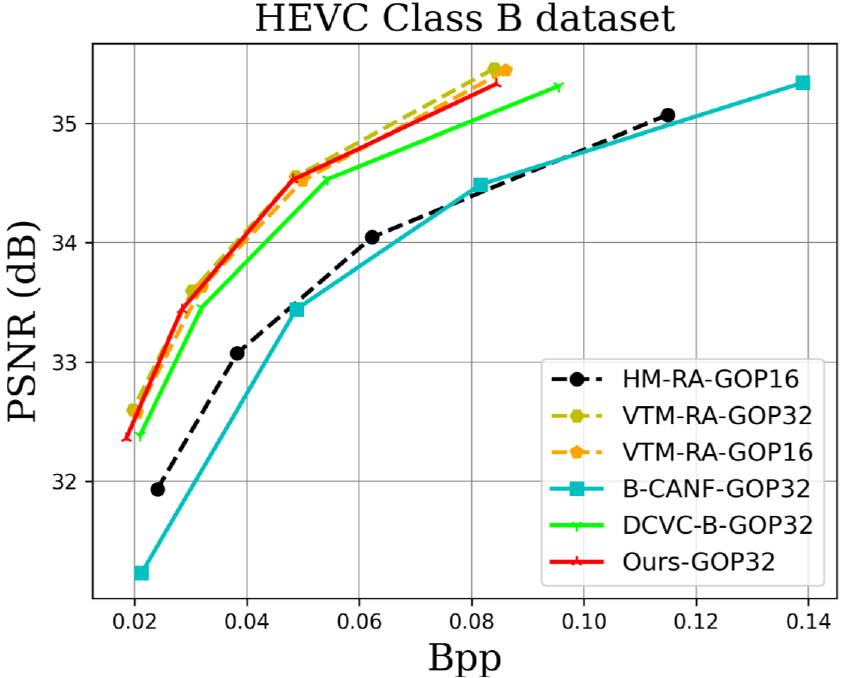}
 \end{minipage}%

  \begin{minipage}[c]{0.31\linewidth}
  \centering
    \includegraphics[width=\linewidth]{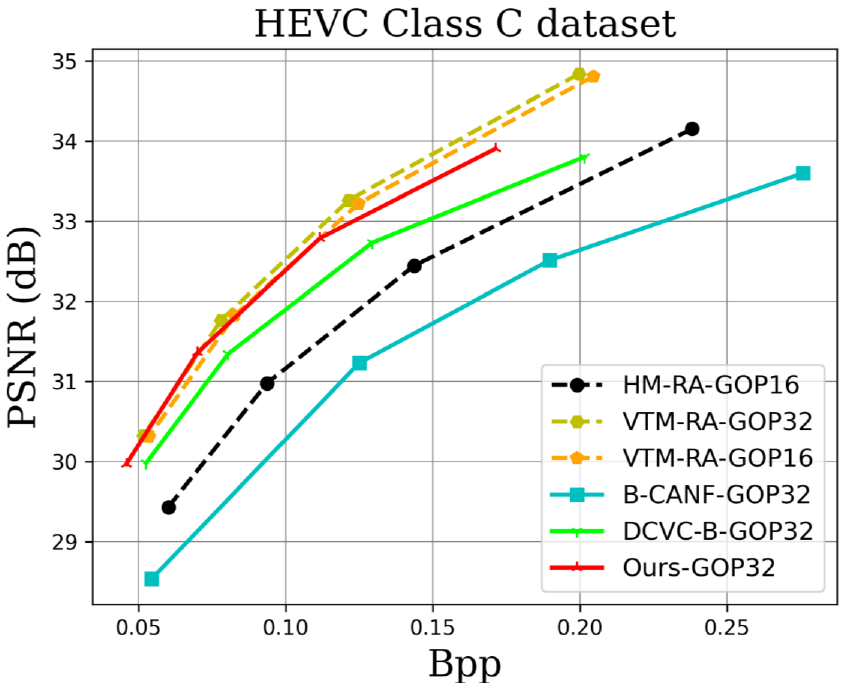}
  \end{minipage}%
  \begin{minipage}[c]{0.31\linewidth}
  \centering
    \includegraphics[width=\linewidth]{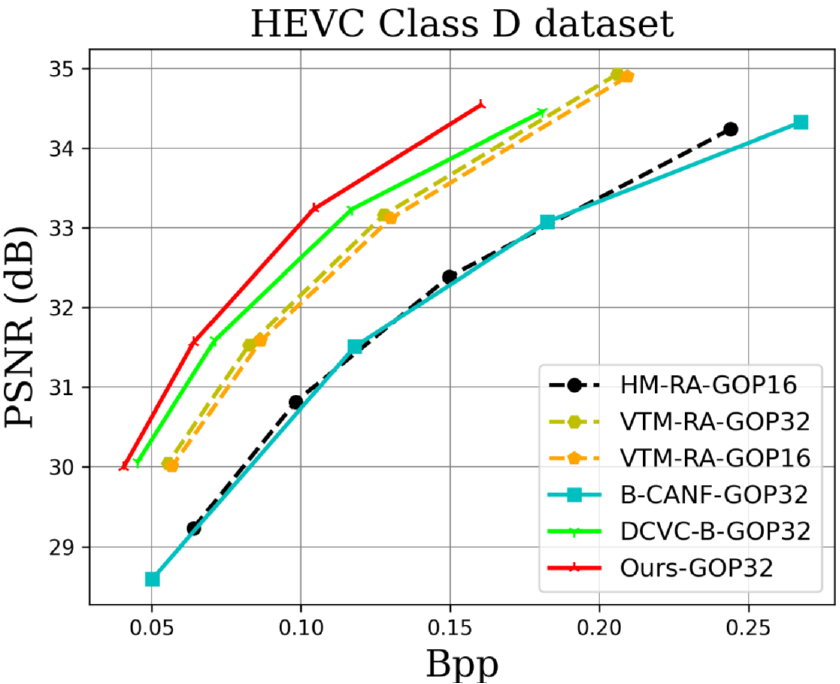}
  \end{minipage}%
  \begin{minipage}[c]{0.31\linewidth}
  \centering
    \includegraphics[width=\linewidth]{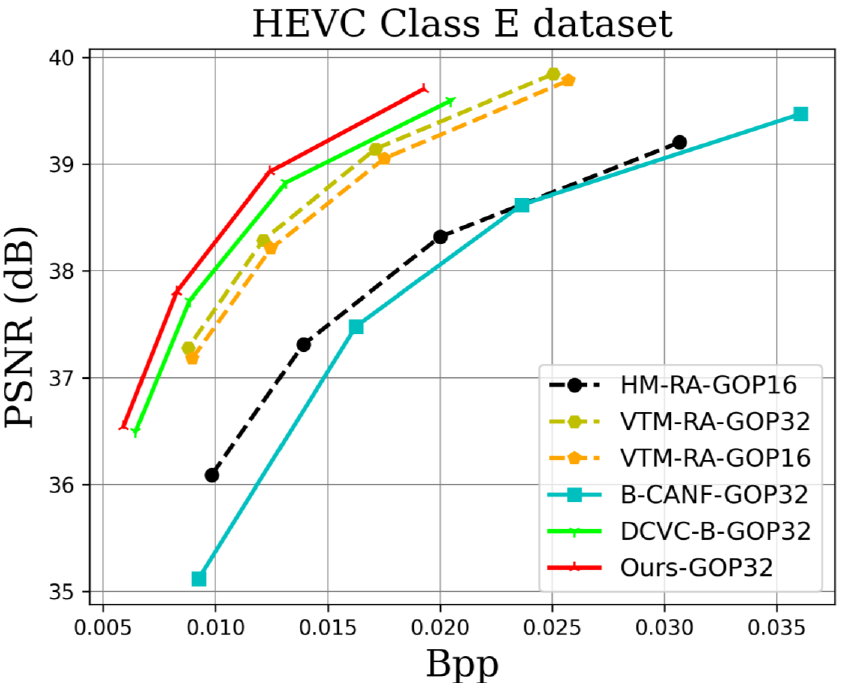}
  \end{minipage}%

    \caption{RD curves of various codecs on MCL-JCV, UVG, and HEVC datasets, measured by PSNR over 97 testing frames.}
  \label{fig:psnr_results_97}
\end{figure*}
\section{Experiments}\label{sec:experiments} 
\subsection{Experimental Setup}
\subsubsection{Training Datasets}
Following previous works~\cite{li2021deep, li2022hybrid,li2023neural,li2024neural,sheng2025bi,nguyenquang2025fast,ho2022canf,chen2023b,sheng2025prediction,sheng2024spatial}, for training, we employ the training split of the Vimeo-90k dataset~\cite{xue2019video}, which comprises 64610 video clips at 448$\times$256 resolution. Since each clip contains only 7 consecutive frames, we utilize this dataset for training stages requiring sequence lengths of 3, 5, and 7 frames~\cite{sheng2025bi}. For longer sequence requirements (17 and 33 frames), we utilize the extended Vimeo collection introduced in~\cite{sheng2025bi}, which offers 9000 higher-resolution (640$\times$360) sequences with 33 frames per clip. The detailed training strategy can be referred to~\cite{sheng2025bi}.

\subsubsection{Evaluation Datasets}
The compression performance is evaluated on three standard benchmark datasets: the MCL-JCV dataset~\cite{wang2016mcl} containing 30 diverse 1920$\times$1080 sequences, the UVG dataset~\cite{mercat2020uvg} with 7 challenging 1080p large-motion sequences, and the HEVC common test sequences~\cite{bossen2013common} comprising 16 videos (Classes B, C, D, E) at resolutions spanning from 416$\times$240 to 1920$\times$1080. We compress 96 (two complete GOPs and one incomplete GOP) and 97 (three complete GOPs) consecutive frames of each test sequence to verify the generalization of our codec for complete GOP and incomplete GOP. The detailed GOP structures can be referred to~\cite{sheng2025bi}.

\subsubsection{Implementation Details}
During training, the Lagrange multiplier $\lambda$ in the loss function is configured as (85, 170, 380, 840) to measure the relative importance of distortion and bit rate. At inference time, our codec achieves different rate points by adjusting its learned quantization steps. The implementation is built in PyTorch with CUDA acceleration, trained using the AdamW optimizer~\cite{kingma2014adam} with default hyperparameter settings. The batch size is set to 8.
The learning rate is gradually reduced from an initial value of $1 \times 10^{-4}$ to a final value of $5 \times 10^{-6}$ throughout the training process.

\subsubsection{Evaluation Metrics}
Following previous works~\cite{li2021deep, li2022hybrid,li2023neural,li2024neural,sheng2025bi,nguyenquang2025fast,ho2022canf,chen2023b,sheng2025prediction,sheng2024spatial}, we compute Peak Signal-to-Noise Ratio (PSNR) in RGB color space for quality assessment, and use bits per pixel (bpp) to measure the bitrate cost.  For comparative analysis between different codecs, we utilize the Bjøntegaard-Delta rate (BD-rate) metric, which calculates the average bitrate difference (\%) relative to an anchor codec. Negative BD-rate values indicate bitrate savings, positive values denote increased bitrate requirements.

\subsection{Quantitative Evaluation Results}
\subsubsection{Comparison with Existing Neural Video Codecs}
To benchmark the compression performance of our proposed codec, we first conduct a comprehensive comparison with state-of-the-art neural video codecs, including both uni-directional inter-frame prediction-based P-frame codecs (DCVC-DC~\cite{li2022hybrid} and DCVC-FM~\cite{li2024neural}) and bi-directional prediction-based B-frame codecs (B-CANF~\cite{chen2023b} and DCVC-B), with all codecs configured using a 32-frame intra-period. When testing 96 frames, our rate-distortion curves presented in Fig.~\ref{fig:psnr_results} demonstrate consistent superiority over competing neural video codecs. Quantitative results in Table~\ref{table:ip32_psnr} reveal that our codec achieves an average bitrate reduction of 35.1\% compared to the HM-RA-GOP16 anchor, significantly outperforming both leading P-frame codecs (DCVC-DC: 24.4\%, DCVC-FM: 24.3\%) and B-frame codec DCVC-B (26.1\%). Notably, on the MCL-JCV dataset, our codec provides more than 10\% compression performance gain over DCVC-B. Similar superiority is achieved when testing 97 frames, as presented in Fig.~\ref{fig:psnr_results_97} and Table~\ref{table:ip32_pnsr_97}, establishing new state-of-the-art neural B-frame coding performance.  \par
It is worth noting that while our method establishes a new state-of-the-art in terms of compression performance, we recognize the distinct advantages of other innovative approaches. For instance, the motion-free concept explored in~\cite{nguyen2024motion} and the fast online motion resolution adaptation in~\cite{nguyenquang2025fast} present compelling alternative pathways. These ideas, focusing on complexity reduction and operational flexibility, offer valuable insights for future work aimed at achieving a better trade-off between performance, complexity, and practicality in neural video coding.\par 
\begin{table*}[t]
\caption{Comparison of BD-rate (\%) against HM-RA-GOP16 using PSNR for reconstruction quality over 97 testing frames.} 
  \centering
\scalebox{0.9}{
\begin{threeparttable}
\begin{tabular}{l|c|c|c|c|c|c|c}
\toprule[1.5pt]
                        &MCL-JCV  &UVG   & HEVC Class B  & HEVC Class C  &HEVC Class D &HEVC Class E &Average\\ \hline
HM-RA-GOP16             &0.0      &0.0   &0.0            &0.0            &0.0          &0.0          &0.0            \\ \hline
VTM-RA-GOP32            &\textbf{--38.4} &\textbf{--35.8}&\textbf{--38.5}         &\textbf{--33.6}         &--30.3       &--37.3       &--35.7         \\ \hline
VTM-RA-GOP16            &--35.8       &--32.4&--35.2         &--31.2         &--28.3       &--33.5       &--32.7         \\ \hline
B-CANF-GOP32            &18.9     &14.0  &7.0            &24.3           &--0.7        &10.2          &--12.3         \\ \hline
DCVC-B-GOP32            &--20.4   &--22.5&--31.1         &--19.7         &--39.2       &--45.4       &--29.7         \\ \hline
Ours-GOP32              &--30.5   &--31.4&\textbf{--38.5}         &--31.1         &\textbf{--45.4}       &\textbf{--50.8}       &\textbf{--38.0}         \\
\bottomrule[1.5pt]
\end{tabular}
\end{threeparttable}}
\label{table:ip32_pnsr_97}
\end{table*}
\subsubsection{Comparison with Traditional Video Codecs}
We further compare with traditional video codecs. Specifically, we include HM-16.20~\cite{HM} (H.265/HEVC reference software) tested under random-access (\texttt{encoder\_randomaccess\_main\_rext}, HM-RA-GOP16) and low-delay (\texttt{encoder\_lowdelay\_vtm}, HM-LDB-GOP4) configurations, along with VTM-13.2 (H.266/VVC reference software) evaluated under random-access (\texttt{encoder\_randomaccess\_vtm\_gop16}, VTM-RA-GOP16 and \texttt{encoder\_randomaccess\_vtm}, VTM-RA-GOP32) and low-delay (\texttt{encoder\_lowdelay\_vtm}, VTM-LDB-GOP8) configurations. As the PSNR is calculated in RGB colorspace, we follow the testing pipelines of~\cite{li2021deep, li2022hybrid,li2023neural,li2024neural,sheng2025bi,nguyenquang2025fast,ho2022canf,chen2023b,sheng2025prediction,sheng2024spatial}, and convert RGB video frames to YUV444 sequences. For better compression performance of traditional codecs, we set their internal colorspace as YUV444.  All codecs are configured with an intra-period of 32. When compressing 96 frames, as reported in Fig.~\ref{fig:psnr_results} and Table~\ref{table:ip32_psnr}, our codec achieves comparable performance to VTM, with an average BD-rate reduction of --34.5\% versus --34.1\% (VTM-RA-GOP16) and --35.8\% (VTM-RA-GOP32). When testing 97 frames, our codec even achieves higher performance than VTM, with an average BD-rate reduction of --38.0\% versus --32.7\% (VTM-RA-GOP16) and --35.7\% (VTM-RA-GOP32). These results demonstrate the strong potential of neural B-frame codecs. \par
It is important to acknowledge that while our results highlight the potential of neural video coding, the field stands on the shoulders of decades of progress in traditional video coding. The sophisticated tools and profound insights developed for standards like H.264/AVC, H.265/HEVC, and H.266/VVC, such as advanced motion estimation methods~\cite{kim2012zoom}, represent a vast repository of knowledge. These well-established principles continue to offer invaluable inspiration and serve as a crucial benchmark for guiding the future development of more efficient, robust, and practical neural video codecs.

\begin{figure*}[t]
  \centering
   \includegraphics[width=0.86\linewidth]{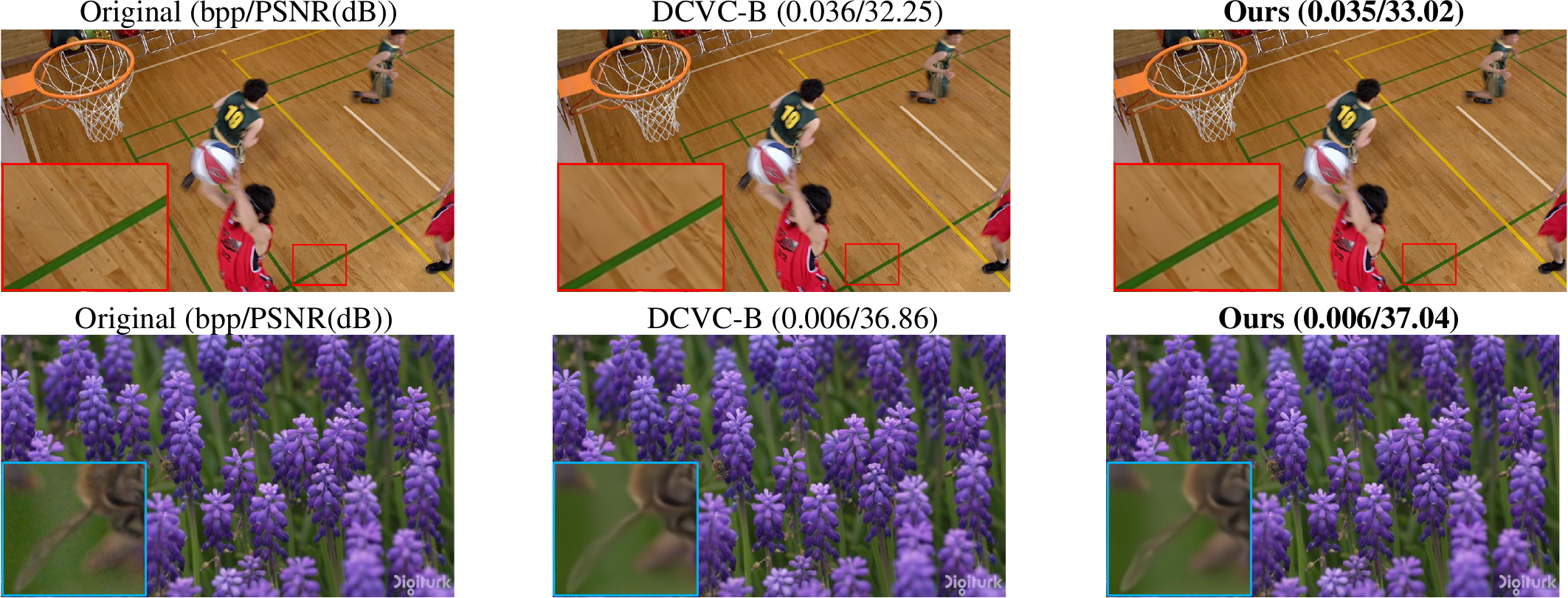}
      \caption{Qualitative comparison for the $17^{th}$ frame of \emph{BasketballDrill\_832x480\_50} and the $2^{nd}$ frame of \emph{HoneyBee\_1920x1080\_120fps\_420\_8bit\_YUV}.}
   \label{fig:subjective}
\end{figure*}
\begin{table}[t]
 \centering
 \caption{Runtime and computational complexity comparison between different codecs. Traditional video codecs are run on a CPU, while neural video codecs are run on a GPU.}
\scalebox{1}{
\begin{tabular}{c|c|c|c|c}
\toprule[1.5pt]
Codecs  & MACs/pixel & Parameter & Enc Time & Dec Time  \\ \hline
HM-RA   &---&---& 92.24s& 0.29s \\ \hline
VTM-RA  &----&---& 1144.70s& 0.37s \\ \hline
DCVC-DC & 1397.90K&18.45M & 0.82s& 0.64s \\ \hline
DCVC-FM &1180.77K &17.02M & 0.74s& 0.53s\\ \hline
B-CANF   &3081.11K &23.66M & 1.49 s& 1.06s \\ \hline
DCVC-B  & 3004.52K& 21.40M & 1.19s& 0.99s  \\ \hline
Ours    & 3382.25K& 26.07M & 1.47s& 1.10s                 \\
\bottomrule[1.5pt]
\end{tabular}}
\label{time}
\end{table}
\subsection{Qualitative Comparison Results}
We also present the qualitative comparison results of our codec and DCVC-B in Fig.~\ref{fig:subjective}. We can observe that the wooden floor textures of our decoded \emph{BasketballDrill\_832x480\_50} sequence maintain richer granular details. The bee's wings of our decoded \emph{HoneyBee\_1920x1080\_120fps\_420\_8bit\_YUV} sequence are reconstructed with sharper transparency and vein patterns. The results indicate that our codec can achieve better visual enhancements while maintaining competitive bitrates.

\subsection{Runtime and Computational Complexity Comparison}
Table~\ref{time} presents a comprehensive comparison of computational complexity and encoding/decoding time across codecs for a 1080p video frame. Traditional codecs (HM, VTM) were evaluated on an Intel(R) Xeon(R) Gold 5118 CPU, while neural codecs were tested on an NVIDIA RTX 3090 GPU. Compared to DCVC-B, our method exhibits increased computational cost (3004.52K $\rightarrow$ 3382.25K MACs/pixel) and model size (21.40M $\rightarrow$ 26.07M parameters), with encoding/decoding time rising from 1.19s $\rightarrow$ 1.47s and 0.99s $\rightarrow$ 1.10s per frame, respectively. While these increases enable superior rate-distortion performance over DCVC-B, further complexity optimization is required. Reducing computational overhead remains a critical focus for our future work.
 \begin{table}[t]
\caption{Ablation Studies to explore the Effectiveness of Different Proposed Methods.}
\centering
\scalebox{0.82}{
\begin{tabular}{c|c|c|c|c|c|c|c}
\toprule[1.5pt]
Model & IDMAE & PBAQS &IMEM &BWF &HIA &BD-rate & $\Delta$ MACs/pixel\\ \hline
$M_1$   & \XSolidBrush  &\XSolidBrush    &\XSolidBrush &\XSolidBrush &\XSolidBrush & 0.0 & 0.0 \\ \hline
$M_2$   &\Checkmark    &\XSolidBrush    &\XSolidBrush &\XSolidBrush &\XSolidBrush & --2.6 & 311.93K\\ \hline
$M_3$   &\Checkmark    &\Checkmark      &\XSolidBrush &\XSolidBrush &\XSolidBrush &--4.9  & 311.93K  \\\hline
$M_4$  & \Checkmark    &\Checkmark      &\Checkmark   &\XSolidBrush &\XSolidBrush &--6.3  & 331.88K  \\\hline
$M_5$  &\Checkmark    &\Checkmark      &\Checkmark   &\Checkmark   &\XSolidBrush &--9.1   & 357.80K \\\hline
$M_6$ &\Checkmark    &\Checkmark      &\Checkmark   &\Checkmark   &\Checkmark   &--10.4  & 377.73K  \\
\bottomrule[1.5pt]
\end{tabular}
}
\label{effectiveness}
\end{table}

\subsection{Ablation Studies}
\subsubsection{Effectiveness of Fine-Grained Motion Compression Method}
\begin{figure}[t]
  \centering
  \begin{minipage}[c]{\linewidth}
  \centering
    \includegraphics[width=\linewidth]{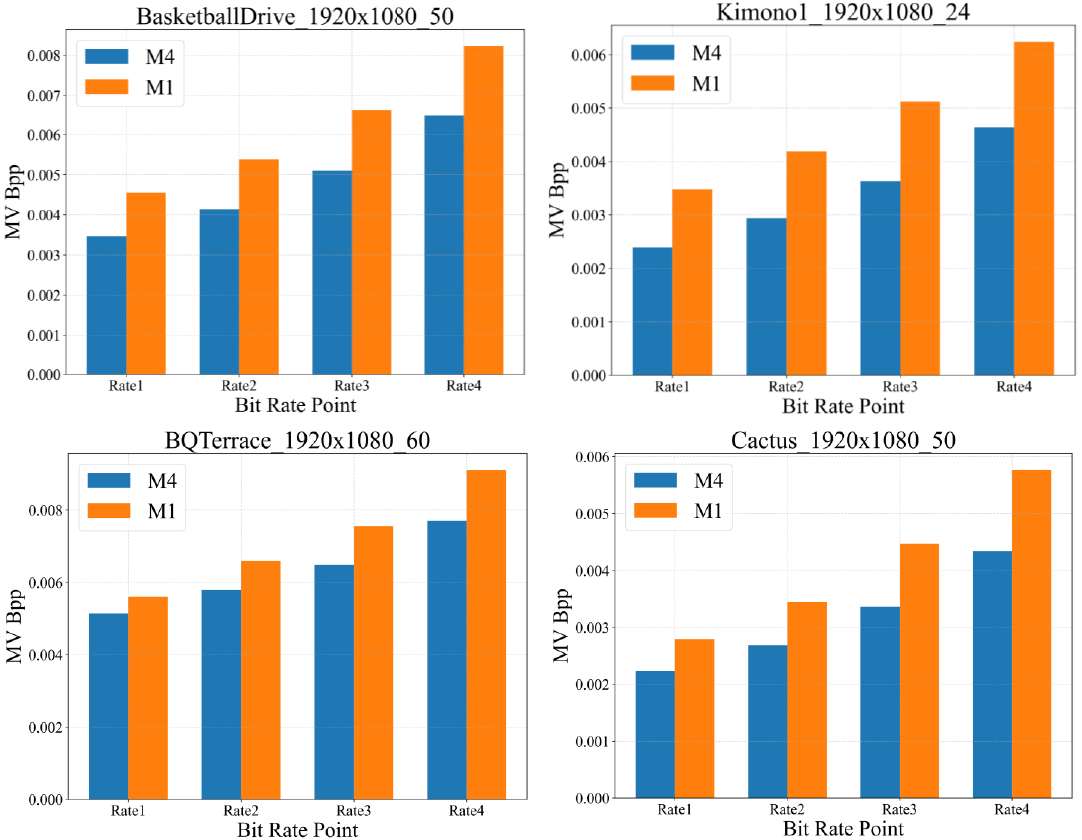}
 \end{minipage}%
    \caption{Comparison of motion bitrate of the models with and without fine-grained motion compression method.}
  \label{fig:ablation_BWA}
\end{figure}

We first conduct an ablation study to validate the effectiveness of our proposed fine-grained motion compression method. To reduce training time, we only use 7-frame video sequences to train the models in the ablation studies, instead of fine-tuning them with the long-sequence dataset~\cite{sheng2025bi}. Our ablation study shows substantial performance gains through progressively integrating the key components into the baseline model $M_1$, as evidenced in Table~\ref{effectiveness}. The interactive dual-branch motion auto-encoder (IDMAE, $M_2$) forms the foundation with a 2.6\% BD-rate improvement, while the per-branch adaptive quantization steps (PBAQS, $M_3$) further deliver a 2.3\% gain (from --2.6\% to --4.9\% BD-rate) by enabling fine-grained rate allocation tailored to each motion direction. The interactive motion entropy model (IMEM, $M_4$) further brings another 1.4\% gain (from --4.9\% to --6.3\% BD-rate), which effectively captures the inherent correlations between bi-directional motion latent representations for enhanced compression efficiency. We compare the motion coding costs of the model with and without our proposed method in Fig.~\ref{fig:ablation_BWA}. We can observe that our method can effectively reduce the bi-directional motion compression bitrate. \par

Regarding the computational complexities of these methods, it is important to note that IDMAE, PBAQS, and IMEM are inherently coupled components of our fine-grained motion compression method. The IDMAE method ($M_2$) establishes the fundamental dual-branch structure necessary for processing bi-directional motions separately, which naturally introduces additional 311.93K MACs/pixel computational complexity increase compared to the single-branch baseline ($M_1$). Without this foundational structure, the subsequent PBAQS and IMEM components would not be feasible. The PBAQS method ($M_3$) maintains the same computational complexity as $M_2$ but increases the model parameters slightly by employing independent quantization steps for each branch instead of the parameter-shared steps used in $M_2$. The IMEM method ($M_4$) adds 19.95K MACs/pixel by incorporating cross-directional dependencies in the motion entropy modeling.\par
\begin{figure}[t]
  \centering
  \begin{minipage}[c]{\linewidth}
  \centering
    \includegraphics[width=\linewidth]{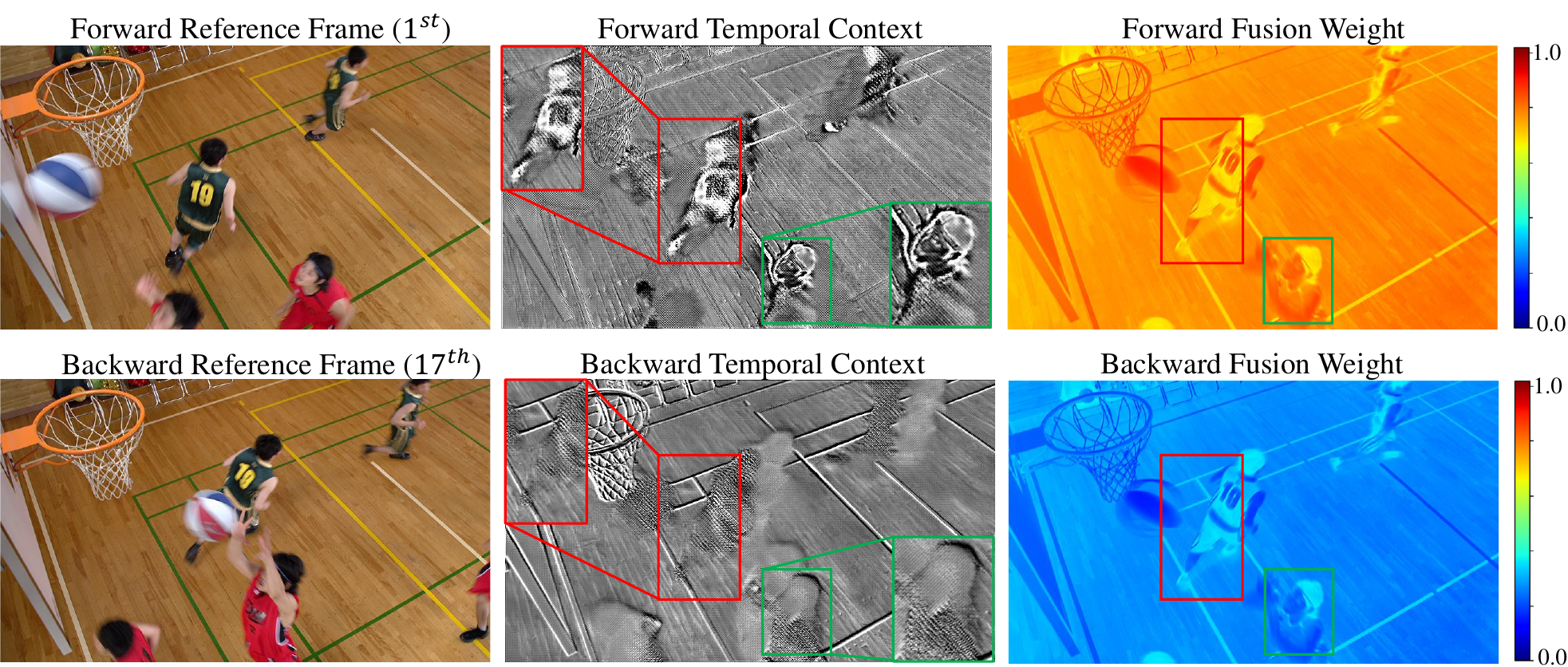}
 \end{minipage}%
    \caption{Visualization of the bi-directional temporal contexts and their corresponding bi-directional fusion weights of the $8^{th}$ frame of \emph{BasketballDrill\_832x480\_50}. }
  \label{fig:ablation_BWF}
\end{figure}
\begin{figure}[t]
  \centering
  \begin{minipage}[c]{\linewidth}
  \centering
    \includegraphics[width=\linewidth]{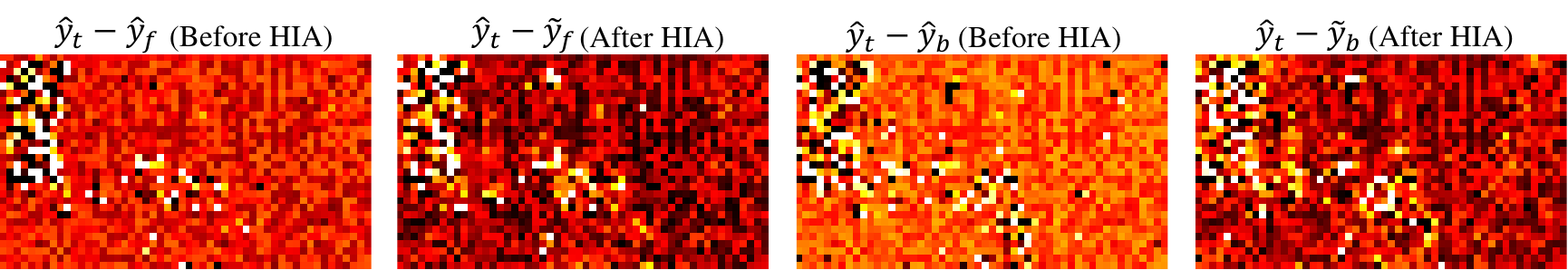}
 \end{minipage}%
    \caption{
    Illustration of the residuals between the contextual latent representation $\hat{y}_t$ of the $6^{th}$ frame of \emph{PartyScene\_832x480\_50} and its bi-directional temporal priors (reference latent representations) before ($\hat{y}_f$, $\hat{y}_b$) and after ($\Tilde{y}_f$, $\Tilde{y}_b$) being proposed by the hyperprior-based implicit alignment. Darker colors indicate smaller residuals, while brighter colors indicate larger residuals.}
  \label{fig:ablation_HIA}
\end{figure}

\subsubsection{Effectiveness of Selective Temporal Fusion Method}
Similarly, we continue to conduct another ablation study to validate the effectiveness of our selective temporal fusion method by progressively integrating its key components into the model $M_4$. As presented in Table~\ref{effectiveness}, the introduction of bi-directional weighting fusion (BWF, $M_5$) first improves the compression performance by 2.8\% BD-rate reduction (from --6.3\% to --9.1\% BD-rate), followed by an additional 1.3\% gain (reaching --10.4\% BD-rate) when incorporating the hyperprior-based implicit alignment (HIA, $M_6$). In terms of computational complexity, the BWF method ($M_5$) introduces additional 25.92K MACs/pixel computational complexity compared to $M_4$, while the HIA method ($M_6$) further adds 19.93K MACs/pixel. The relatively modest increase in complexity shows that our proposed selective temporal fusion method achieves effective performance improvement with limited computational overhead. \par

We further visualize the predicted bi-directional temporal contexts along with their corresponding fusion weights in Fig.~\ref{fig:ablation_BWF}. Taking the $8^{th}$ frame from \emph{BasketballDrill\_832x480\_50} as an instance, we can observe that the forward temporal context exhibits relatively higher prediction quality than the backward temporal context. As a result, our proposed BWF assigns larger fusion weights to the forward temporal context to preserve its temporal information more effectively, while assigning smaller fusion weights to the backward temporal context to mitigate its negative impacts. In addition, we also visualize the residuals between the contextual latent representation $\hat{y}_t$ of the $6^{th}$ frame of \emph{PartyScene\_832x480\_50} and its bi-directional temporal priors before ($\hat{y}_f, \hat{y}_b$) and after ($\Tilde{y}_f, \Tilde{y}_b$) being processed by our proposed HIA. As illustrated in Fig.~\ref{fig:ablation_HIA}, our HIA can effectively reduce the residual magnitudes, indicating a significant decrease in the local misalignment of temporal priors. 

\section{Conclusion and Future Work}\label{sec:conclusion}
We first propose a fine-grained motion compression method for neural B-frame video coding. By leveraging an interactive dual-branch motion auto-encoder,  per-branch adaptive quantization steps, and an interactive motion entropy model, the bi-directional motion coding costs are effectively reduced. Furthermore, we propose a selective temporal fusion method. Using bi-directional fusion weights and hyperprior-based implicit alignment, our method enables discriminative utilization of bi-directional temporal contexts and temporal priors in the contextual encoder-decoder and contextual entropy model. Experimental results verify that our proposed codec outperforms state-of-the-art neural B-frame codecs and achieves comparable or even superior compression performance to the H.266/VVC reference software under random-access configuration. \par

While the proposed codec achieves impressive compression performance, its increased computational complexity relative to existing neural codecs presents a key challenge for practical deployment. Our future work will, therefore, focus on achieving a more favorable complexity-performance trade-off. A promising direction, inspired by motion-free neural B-frame coding approaches, is the development of a hybrid coding framework. Specifically, within a hierarchical B-frame structure, frames at higher temporal layers exhibit smaller motion due to their proximity to the references. For these frames, we plan to leverage lightweight, interpolation-based prediction to generate temporal contexts, effectively bypassing the computationally intensive processes of motion estimation, compression, and compensation. We anticipate that this content-adaptive strategy will significantly reduce the average computational cost while retaining the majority of the compression performance gains achieved by our current design.

\section{Acknowledgment}\label{sec:Acknowledgment}
We would like to express our gratitude to Zhirui Zuo for his assistance in dataset processing.

\bibliographystyle{ieeetr}
\bibliography{ref}
\begin{IEEEbiography}[{\includegraphics[width=1in,height=1.25in,clip,keepaspectratio]{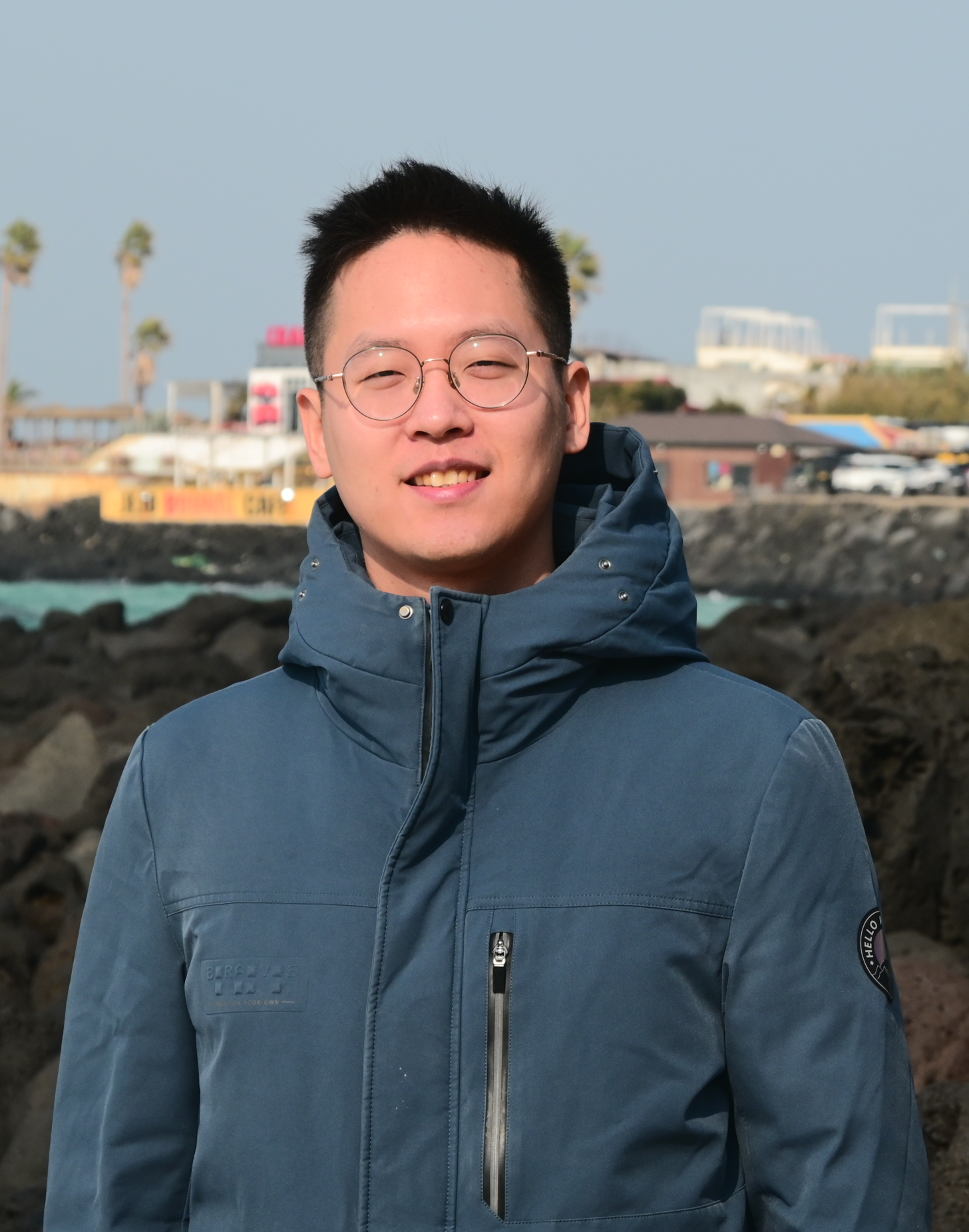}}]{Xihua Sheng} (Member, IEEE) received the B.S. degree in automation from Northeastern University, Shenyang, China, in 2019, and the Ph.D. degree in electronic engineering from University of Science and Technology of China (USTC), Hefei, Anhui, China, in 2024. 
He is currently a Postdoctoral Fellow in computer science from City University of Hong Kong. 
His research interests include image/video/point cloud coding, signal processing, and machine learning.
\end{IEEEbiography}

\begin{IEEEbiography}[{\includegraphics[width=1in,height=1.25in,clip,keepaspectratio]{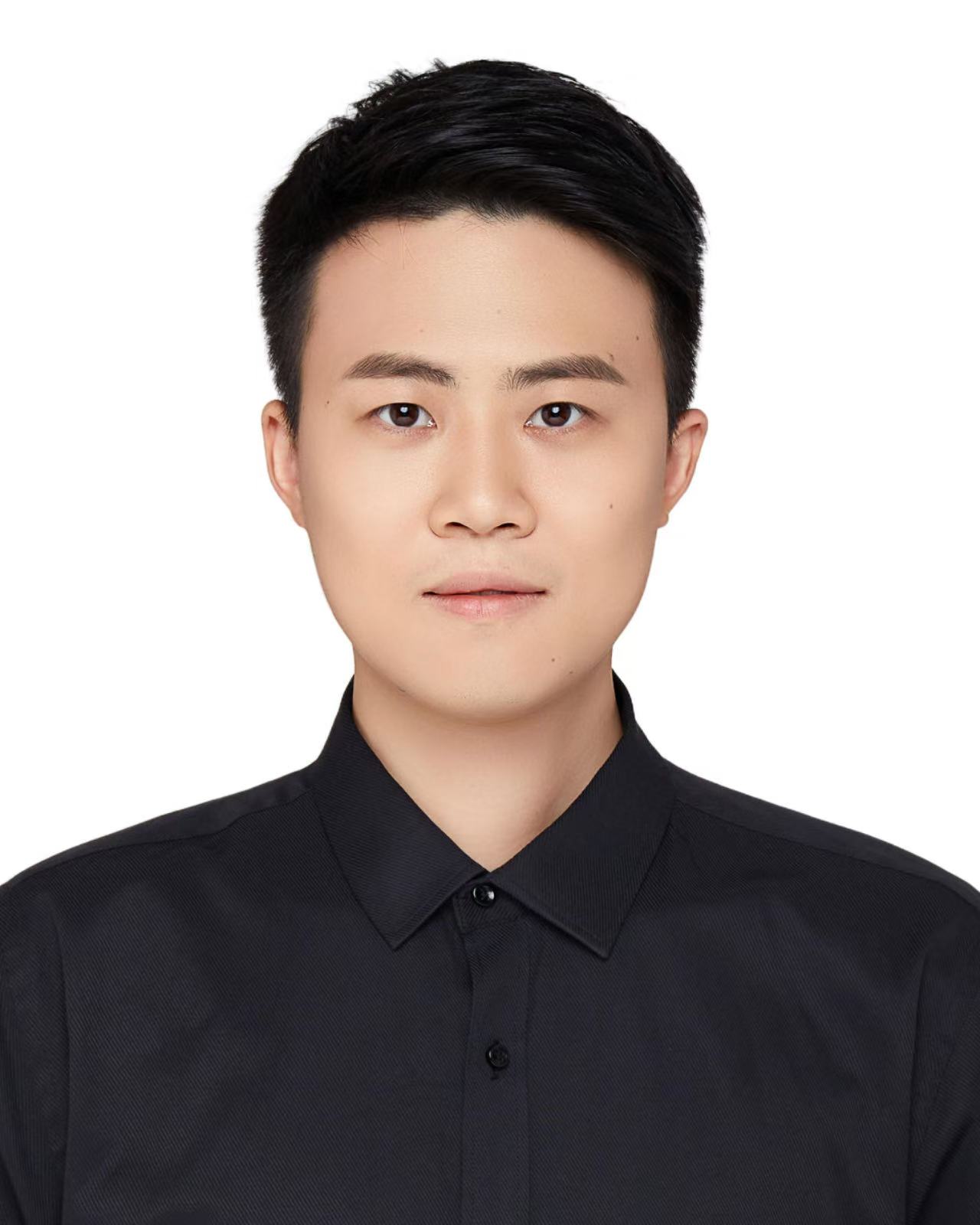}}] {Peilin Chen}
    received the B.S. degree in Software Engineering from Sun Yat-sen University, Guangzhou, China, in 2018 and the Ph.D. degree in computer science from the City University of Hong Kong, Hong Kong SAR, China, in 2023. He is currently a Postdoctoral Researcher with the Department of Computer Science, City University of Hong Kong. His current research interests include visual data processing and semantic communication.
\end{IEEEbiography}

\begin{IEEEbiography}[{\includegraphics[width=1in,height=1.25in,clip,keepaspectratio]{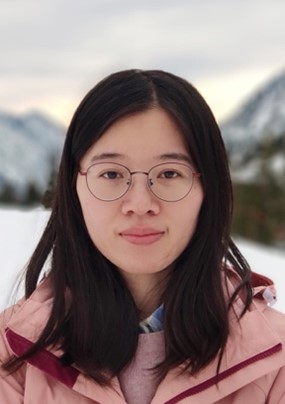}}] {Meng Wang} (Member, IEEE) received the B.S. degree in electronic information engineering of Honors Program from China Agricultural University, Beijing, China, in 2015, the M.S. degree in computer application technology from Peking University, Beijing, China, in 2018, and the Ph.D. degree in computer science from the City University of Hong Kong, Hong Kong, China, in 2021. She was a postdoc with the Department of Computer Science, City University of Hong Kong. She was a research intern in Bytedance Inc., since 2017. She is currently an Assistant Professor with the School of Data Science, Lingnan University, Hong Kong. She has authored or coauthored more than 50 refereed journal articles/conference papers. She has been actively participated in the development of VVC, AVS3 and JPEG-AI standards. Her research interests include data compression and image/video coding.
\end{IEEEbiography}

\begin{IEEEbiography}[{\includegraphics[width=1in,height=1.25in,clip,keepaspectratio]{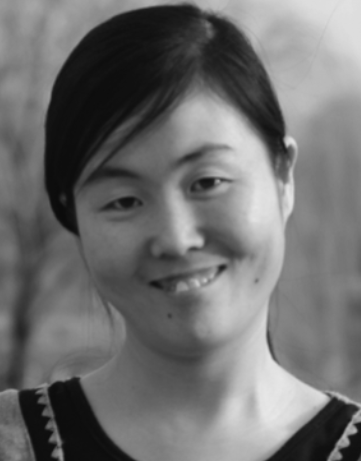}}] {Li Zhang} (Senior Member, IEEE)  received the Ph.D. degree from the Institute of Computing Technology, Chinese Academy of Sciences, in 2009. Currently, she leads the Multimedia Laboratory, ByteDance Inc., pioneering cutting-edge technologies in multimedia. With a focus on video compression, streaming, and signal processing, she holds more than 700 granted U.S. patents and has published more than 100 technical articles in book chapters, journals, and conference proceedings. Additionally, she has made more than 600 adopted standardization contributions to various standards, such as H.266/VVC, H.265/HEVC, AVS, IEEE 1857, H.264/AVC, G-PCC, and JPEG AI. Her research has been recognized with numerous awards, including the Best Paper Award at the 2022 ISCAS Visual Signal Processing and Communications Track and the Top Ten Best Paper Award at the 2021 IEEE PCS. She has also secured several first-place accolades in international challenges and received Certificates of Appreciation for her exceptional contributions to the IEEE 1857 Standard in 2013 and 2021. She has served as an editor, a software coordinator, and the chair for core experiments in standard groups. She has organized and co-chaired multiple special sessions and grand challenges at conferences. She holds the position of the Publicity Subcommittee Chair of the Technical Committee on Visual Signal Processing and Communications in the IEEE CAS Society. She is an Associate Editor of IEEE Transactions on Circuits and Systems for Video Technology.
\end{IEEEbiography}

\begin{IEEEbiography}[{\includegraphics[width=1in,height=1.25in,clip,keepaspectratio]{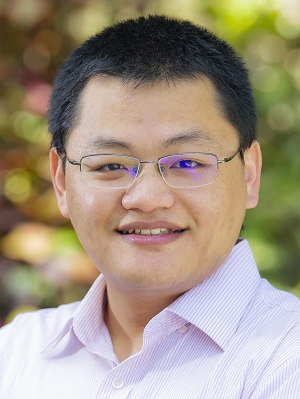}}] {Shiqi Wang} (Senior Member, IEEE) received the PhD degree in computer application technology from Peking University, in 2014. He is currently a Professor with the Department of Computer Science, City University of Hong Kong, Hong Kong. He has proposed more than 70 technical proposals to ISO/MPEG, ITUT, and AVS standards. He authored or coauthored more than 300 refereed journal articles/conference papers, including more than 100 IEEE Transactions. His research interests include semantic and visual communication, AI generated content management, machine learning, information forensics and security, and image/video quality assessment. He received the Best Paper Award from IEEE VCIP 2019, ICME 2019, IEEE Multimedia 2018, and PCM 2017. His coauthored article received the Best Student Paper Award in the IEEE ICIP 2018. He was the TPC Chair of ICME 2024. He served or serves as an associate editor for IEEE Transactions on Circuits and Systems for Video Technology, IEEE Transactions on Multimedia, IEEE Transactions on Image Processing, and IEEE Transactions on Cybernetics.
\end{IEEEbiography}

\begin{IEEEbiography}[{\includegraphics[width=1in,height=1.25in,clip,keepaspectratio]{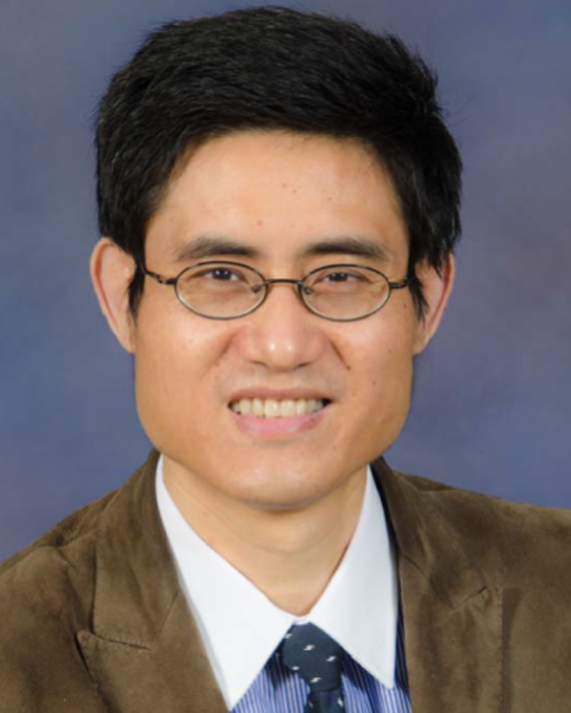}}] {Dapeng Oliver Wu} (Fellow, IEEE)   received a B.E. degree in electrical engineering from Huazhong University of Science and Technology, Wuhan, China, in 1990, an M.E. degree in electrical engineering from Beijing  University of Posts and Telecommunications, Beijing, China, in 1997, and a Ph.D. degree in electrical and computer engineering from Carnegie Mellon University, Pittsburgh, PA, in 2003.\par
He is Yeung Kin Man Chair Professor of Network Science, and Chair Professor of Data Engineering at the Department of Computer Science, City University of Hong Kong. Previously, he was on the faculty of University of Florida, Gainesville, FL, USA and was the director of NSF Center for Big Learning, USA. His research interests are in the areas of artificial intelligence, network science, communications, signal processing, computer vision, and biomedical engineering. He received University of Florida Term Professorship Award in 2017, University of Florida Research Foundation Professorship Award in 2009, AFOSR Young Investigator Program (YIP) Award in 2009, ONR Young Investigator Program (YIP) Award in 2008, NSF CAREER award in 2007, the IEEE Transactions on Emerging Topics in Computational Intelligence (TETCI) Outstanding Paper Award for Year 2025, the IEEE Circuits and Systems for Video Technology (CSVT) Transactions Best Paper Award for Year 2001, and the Best Paper Awards in IEEE GLOBECOM 2011 and International Conference on Quality of Service in Heterogeneous  Wired/Wireless Networks (QShine) 2006.\par
He has served as founding Editor in Chief of Transactions of Artificial Intelligence, Editor in Chief of IEEE Transactions on Network Science and Engineering, founding Editor in Chief of Journal of Advances in Multimedia, Editor-at-Large for IEEE Open Journal of the Communications Society, and Associate Editor for IEEE Transactions on Cloud Computing, IEEE Transactions on Communications, IEEE Transactions on Signal and
 Information Processing over Networks, IEEE Signal Processing Magazine, IEEE Transactions on Circuits and Systems for Video Technology, IEEE Transactions on Wireless Communications and IEEE Transactions on Vehicular Technology. He has served as Technical Program Committee (TPC) Chair for IEEE INFOCOM 2012, and TPC chair for IEEE International Conference on Communications (ICC 2008), Signal Processing for Com
munications Symposium, and as a member of executive committee and/or technical program committee of over 100 conferences. He was elected as a Distinguished Lecturer by IEEE Vehicular Technology Society in 2016. He is an IEEE Fellow.
\end{IEEEbiography}

\end{document}